\documentclass[twocolumn]{aastex631}

\usepackage{amsmath}
\usepackage{gensymb}
\usepackage{multirow}

\shorttitle{Optical/infrared analysis of GRB 250702B}
\shortauthors{}
\graphicspath{{./}{figures/}}

\definecolor{blazeorange}{rgb}{1.0, 0.4, 0.0}
\definecolor{seagreen}{rgb}{0.18, 0.55, 0.34}
\definecolor{darkgreen}{rgb}{0.08, 0.45, 0.2}
\definecolor{rufous}{rgb}{0.66, 0.11, 0.03}
\definecolor{royalfuchsia}{rgb}{0.79, 0.17, 0.57}
\definecolor{scarlet}{rgb}{1.0, 0.13, 0.0}
\definecolor{royalpurple}{rgb}{0.47, 0.32, 0.66}

\begin{document}

\title{Optical/infrared observations of the extraordinary GRB 250702B: \\a highly obscured afterglow in a massive galaxy consistent with multiple possible progenitors}

\newcommand{\unc}{
    Department of Physics and Astronomy, 
    University of North Carolina at Chapel Hill, 
    Chapel Hill, NC 27599-3255, USA
}
\newcommand{\cmu}{
    McWilliams Center for Cosmology and Astrophysics, Department of Physics, Carnegie Mellon University, Pittsburgh, PA 15213, USA
}
\newcommand{\manchester}{
    Jodrell Bank Centre for Astrophysics, Alan Turing Building, University of Manchester, Oxford Road, Manchester M13 9PL, UK
}
\newcommand{\caltech}{Cahill Center for Astrophysics, California Institute of Technology, MC 249-17, 1216 E California Boulevard, Pasadena, CA, 91125, USA}
\newcommand{\grb}{GRB\,250702B}

\author[0000-0001-8544-584X]{Jonathan Carney}
    \affiliation{\unc}
    \email{jcarney@unc.edu}
\author[0000-0002-8977-1498]{Igor Andreoni}
    \affiliation{\unc}
\author[0000-0002-9700-0036]{Brendan O'Connor}
    \altaffiliation{McWilliams Fellow}
    \affiliation{\cmu}
\author[0009-0006-7990-0547]{James Freeburn}
    \affiliation{\unc}
\author[0000-0003-0516-3485]{Hannah Skobe}
    \affiliation{\cmu}
\author[0009-0008-8642-5275]{Lewi Westcott}
    \affiliation{\manchester}
\author[0009-0001-0574-2332]{Malte Busmann}
    \affiliation{University Observatory, Faculty of Physics, Ludwig-Maximilians-Universität München, Scheinerstr. 1, 81679 Munich, Germany}
\author[0000-0002-6011-0530]{Antonella Palmese}
    \affiliation{\cmu}
\author[0000-0002-9364-5419]{Xander J. Hall}
    \affiliation{\cmu}
\author[0000-0003-0516-2968]{Ramandeep Gill}
    \affiliation{Instituto de Radioastronom\'ia y Astrof\'isica, Universidad Nacional Aut\'onoma de M\'exico, Antigua Carretera a P\'atzcuaro $\#$ 8701,  Ex-Hda. San Jos\'e de la Huerta, Morelia, Michoac\'an, C.P. 58089, M\'exico }
    \affiliation{Astrophysics Research Center of the Open university (ARCO), The Open University of Israel, P.O Box 808, Ra’anana 4353701, Israel}

\author[0000-0001-7833-1043]{Paz Beniamini}
    \affiliation{Department of Natural Sciences, The Open University of Israel, P.O Box 808, Ra'anana 4353701, Israel}
    \affiliation{Astrophysics Research Center of the Open University (ARCO), The Open University of Israel, P.O Box 808, Ra'anana 4353701, Israel}
    \affiliation{Department of Physics, The George Washington University, 725 21st Street NW, Washington, DC 20052, USA}

\author[0000-0003-3765-6401]{Eric R. Coughlin}
    \affiliation{Department of Physics, Syracuse University, Syracuse, NY 13210, USA,}
    
\author[0000-0002-5740-7747]{Charles D. Kilpatrick}
    \affiliation{Center for Interdisciplinary Exploration and Research in Astrophysics (CIERA), Northwestern University, 1800 Sherman Ave, Evanston, IL 60201, USA}
\author[0000-0002-8935-9882]{Akash Anumarlapudi}
    \affiliation{\unc}
\author[0000-0001-9380-6457]{Nicholas M. Law}
    \affiliation{\unc}
\author[0000-0002-6339-6706]{Hank Corbett}
    \affiliation{\unc}

\author[0000-0002-2184-6430]{Tomas Ahumada}
    \affiliation{\caltech}
\author[0000-0003-0853-6427]{Ping Chen}
    \affiliation{Institute for Advanced Study in Physics, Zhejiang University, Hangzhou 310027, China}
    \affiliation{Institute for Astronomy, School of Physics, Zhejiang University, Hangzhou 310027, China}
\author[0000-0003-1949-7638]{Christopher Conselice}
    \affiliation{\manchester}
\author[0000-0002-2651-7038]{Guillermo Damke}
    \affiliation{Cerro Tololo Inter-American Observatory/NSF NOIRLab, Casilla 603, La Serena, Chile}
\author[0000-0001-8372-997X]{Kaustav K. Das}
    \affiliation{\caltech}
\author[0000-0002-3653-5598]{Avishay Gal-Yam}
\affiliation{Department of Particle Physics and Astrophysics
Weizmann Institute of Science, 234 Herzl St., Rehovot, Israel}
\author[0000-0003-3270-7644]{Daniel Gruen}
    \affiliation{University Observatory, Faculty of Physics, Ludwig-Maximilians-Universität München, Scheinerstr. 1, 81679 Munich, Germany}
    \affiliation{Excellence Cluster ORIGINS, Boltzmannstr. 2, 85748 Garching, Germany}
\author[0000-0002-0856-3663]{Steve Heathcote}
    \affiliation{Cerro Tololo Inter-American Observatory/NSF NOIRLab, Casilla 603, La Serena, Chile}
\author[0000-0001-7201-1938]{Lei Hu}
    \affiliation{\cmu}
\author[0000-0003-2758-159X]{Viraj Karambelkar}
    \affiliation{\caltech}
\author[0000-0002-5619-4938]{Mansi Kasliwal}
    \affiliation{\caltech}
\author[0000-0002-6633-7891]{Kathleen Labrie}
    \affiliation{Gemini Observatory/NSF's NOIRLab, 670 N. A'ohoku Place, Hilo, HI 96720, USA}
    \author[0000-0003-1386-7861]{Dheeraj Pasham}
	\affiliation{Eureka Scientific, 2452 Delmer Street Suite 100, Oakland, CA 94602-3017, USA}
    \affiliation{Department of Physics, The George Washington University, Washington, DC 20052, USA}
\author[0000-0002-5466-3892]{Arno Riffeser}
    \affiliation{University Observatory, Faculty of Physics, Ludwig-Maximilians-Universität München, Scheinerstr. 1, 81679 Munich, Germany}
\author[0009-0003-1323-9774]{Michael Schmidt}
    \affiliation{University Observatory, Faculty of Physics, Ludwig-Maximilians-Universität München, Scheinerstr. 1, 81679 Munich, Germany}
\author[0000-0002-4477-3625]{Kritti Sharma}
    \affiliation{\caltech}
\author{Silona Wilke}
    \affiliation{University Observatory, Faculty of Physics, Ludwig-Maximilians-Universität München, Scheinerstr. 1, 81679 Munich, Germany}

\author[0000-0001-6000-3463]{Weicheng Zang}
\affiliation{Center for Astrophysics $|$ Harvard \& Smithsonian, 60 Garden St., Cambridge, MA 02138, USA}

\begin{abstract} 

GRB 250702B was the longest gamma-ray burst ever detected, with a duration that challenges standard collapsar models and suggests an exotic progenitor. 
We collected a rich set of optical and infrared follow-up observations of its rapidly fading afterglow using a suite of telescopes including the W. M. Keck Observatory, the Gemini telescopes, the Magellan Baade Telescope, the Victor M. Blanco 4-meter telescope, and the Fraunhofer Telescope at Wendelstein Observatory.
Our analysis reveals that the afterglow emission is well described by forward shock emission from a highly obscured relativistic jet. 
Deep photometric observations of the host galaxy reveal a massive ($10^{10.66} M_\odot$), dusty, and extremely asymmetric system that is consistent with two galaxies undergoing a major merger.
The galactocentric offset, host galaxy properties, and jet characteristics disfavor a jetted TDE around a supermassive black hole but do not definitively distinguish between competing progenitor scenarios.
We find that the afterglow and host are consistent with a range of progenitors including an atypical collapsar, a merger between a helium star and a stellar mass black hole, the disruption of a star by a stellar mass compact object (micro-TDE), and the tidal disruption of a star by an off-nuclear intermediate mass black hole.

\end{abstract}


\section{Introduction}
\label{sec:intro}

Relativistic jets can illuminate the entire electromagnetic spectrum, reaching very high luminosities from gamma-rays to the radio when the observer's line of sight is fortuitously aligned with the jet axis. Among the plethora of high-energy transients, gamma-ray bursts (GRBs) are the most common. Progenitors of the majority of GRBs include collapsars and compact object mergers \citep[see e.g.][for a review]{Piran2004}. The prompt emission from GRBs is usually found to last from $\sim 0.01$\,s up to several minutes \citep{Kouveliotou1993}, but a rare population of so-called ultralong GRBs exists \citep{Levan2014ultralong}, where prompt emission can last for several thousand seconds \citep[see also][]{Tikhomirova2005}. The progenitors of these systems are believed to be massive stars with particularly large radii, which is supported by the emergence of bright supernovae after some ultralong GRBs \citep[e.g.,][]{Stratta2013, Levan2014ultralong}. An helium-star merger progenitor \citep{Fryer1998} has also been proposed, for instance in the case of ultralong GRB~101225A \citep{Thone2011}.

Although exceptionally rare, the disruption of a star by a massive black hole may also produce relativistic jets, in which case the gamma-ray activity may last for a few days. This extended emission from a relativistic tidal disruption event (TDE) was observed for Sw J1644+57 \citep{Bloom2011Sci, burrow11, Levan2011Sci, Zauderer2011Nat}, Sw J2058+05 \citep{Cenko2012, Pasham2015}, Sw J1112-82 \citep{brown15, Brown2017}, and AT2022cmc \citep{Andreoni2022, Pasham2023}. The high-energy spectrum of these sources is softer than typical GRBs, thus the emission is long-lasting predominantly in the hard X--rays.

On 2025 July 2, the Gamma-ray Burst Monitor (GBM) onboard the {\it Fermi} satellite detected multiple GRBs likely originating from the same source \citep{GRB250702B_overlap_gcn}. They were observed between 2025-07-02 13:56:00 UT and 2025-07-02 16:21:33 UT and designated GRB 250702B \citep{GRB250702B_det_gcn}, GRB 250702C \citep[later retracted;][]{GRB250702C_det_gcn, GRB250702C_dissoc_gcn}, GRB 250702D \citep{GRB250702D_det_gcn}, and GRB 250702E \citep{GRB250702E_det_gcn}. Henceforth we will refer to the event as \grb, in accordance with standard GRB nomenclature. The transient was later reported to have been concurrently observed by other facilities including {\it Neil Gehrels Swift Observatory} \citep[hereafter {\it Swift;}][]{Gehrels2004} Burst Array Telescope's Gamma-ray Urgent Archiver for Novel Opportunities (BAT-GUANO) \citep{grb250702B_guano_gcn,OConnor2025}, the \textit{Monitor of All-sky X-ray Image} (\textit{MAXI}) \citep{EP250702a_maxi_gcn}, \textit{Konus-Wind} \citep{EP250702a_konus_gcn}, and the \textit{Space-based multi-band astronomical Variable Objects Monitor} (\textit{SVOM}) \citep{SVOM_epoch_1}. \textit{Konus-Wind} detected multiple over-lapping X-ray emission episodes over hours from 2025-07-02 12:40 UT to 2025-07-02 16:50 UT \citep{EP250702a_konus_gcn}, suggesting an extreme duration for the transient. A complete gamma-ray analysis combining \textit{Konus-Wind} and \textit{Fermi} finds a gamma-ray duration of $\sim 25$\,ks  \citep{Neights2025}.
This extraordinary duration was further extended by retroactive analysis of stacked \textit{Einstein Probe} \citep[EP;][]{Yuan2025} Wide-field X-ray Telescope (WXT) images, which revealed emission as early as July 1st \citep{EP250702a_det_gcn}, 
giving it a central engine duration of $\sim100$\,ks \citep{Neights2025}. 
This makes \grb\ the longest-duration gamma-ray bursts ever detected - the previous record holder GRB 111209A lasted $\sim15$\,ks in the gamma rays and had a central engine duration of $\sim25$\,ks \citep{Gendre2013}. The exceptionally long duration of \grb\ complicates explanation from the known population of GRBs \citep{Piran2004, Woosley2006}. 

The EP mission detected a soft X-ray counterpart (designated EP250702a) at 2025-07-02 02:53 UT, with a localization accuracy of 2.4\, arcminutes. Follow-up observations with the {\it Swift} X-ray Telescope (XRT) further improved the localization to a radius of 2.0\arcsec\ centered on J2000 coordinates RA =18h 58m 45.61s, Dec = $-$07d 52$'$ 26.9$''$ \citep{EP250702a_XRT_GCN}.  
Following these initial reports, ultraviolet (UV) optical and near-infrared (nIR) follow-up observations ensued. No UV or optical counterparts were identified in these initial searches (see Table\,\ref{tab:observations_table}).
At 0.746 days after the first \textit{Fermi} detection, Very Large Telescope (VLT) HAWK-I observations identified a nIR counterpart \citep{Levan_long_GRB} at coordinates RA=18h 58m 45.57s, Dec= $-$07d 52$'$ 26.2$''$.
Subsequent follow-up observations confirmed that the nIR counterpart was rapidly fading and coincident with an extended nIR galaxy \citep{Levan_long_GRB}. Neither the transient nor the host galaxy were initially detected in any optical photometric band during follow-up, suggesting significant extinction along the line of sight. 

In this work, we focus on optical and nIR observations of \grb\ and its host galaxy. The paper is organized as follows. We describe our observations in Sec.\,\ref{sec:observations_short} and in Appendix \ref{sec:observations_long}. Modeling of both the host galaxy and the transient emission is presented in Sec.\,\ref{sec:analysis}. We discuss our results and their implications for the progenitor in Sec.\,\ref{sec:discussion}. We present our conclusions in Sec.\,\ref{sec:conclusion}.

Throughout this work we adopt a standard $\Lambda$CDM cosmology with $H_0$\,$=$\,$67.4$ km s$^{-1}$ Mpc$^{-1}$, $\Omega_\textrm{m}$\,$=$\,$0.315$, and $\Omega_\Lambda$\,$=$\,$0.685$ \citep{Planck2020}. All magnitudes are reported in the AB system.

\section{Observations}
\label{sec:observations_short}
Following the initial \textit{Fermi} and EP detections of \grb\, we began an optical to nIR follow-up campaign, beginning with $griz$ observations using the Dark Energy Camera \citep[DECam;][]{Flaugher2015} mounted on the Victor M. Blanco 4-meter Telescope at Cerro Tololo Inter-American Observatory (Program ID 2025A-729671; PI: Palmese). After the XRT localization and counterpart detection by VLT became available, we continued to monitor the afterglow with the Fraunhofer Telescope at Wendelstein Observatory (FTW) using the Three Channel Imager \citep[3KK;][]{3kk_spie} for simultaneous optical and nIR imaging. We also acquired deep $z$ band imaging with both Gemini Multi Object Spectrographs on Gemini South (GMOS-S; program ID GS-2025A-Q-117; PI: Andreoni) and Gemini North (GMOS-N; program ID GN-2025A-Q-213; PI: Andreoni), and deep nIR monitoring with the Multi-Object Spectrograph for Infrared Exploration \citep[MOSFIRE;][]{mosfire_spie} on the W. M. Keck Observatory telescope I (PI: Kasliwal, Program ID: C348), the NOAO Extremely Wide Field Infrared Imager (NEWFIRM; \citealt{Autry03}) on the Victor M. Blanco 4-meter Telescope at Cerro Tololo Inter-American Observatory, and the FourStar infrared camera \citep{Persson2013} mounted on the 6.5 m Magellan Baade Telescope. Observations extended from $\sim15$ hours to $\sim 41$ days after the initial \textit{Fermi} detection. Details for these observations can be found in Appendix \ref{sec:observations_long}.

To measure the transient flux, we performed image subtraction using our late-time Magellan/FourStar observations as templates for the FTW/3KK, Keck/MOSFIRE, and NEWFIRM images. We also reprocessed publicly available data from the \textit{Hubble Space Telescope} \citep[\textit{HST}, Program ID 17988; PI: Levan, previously published in][]{Levan_long_GRB}. To reduce host contamination in {\it HST photometry}, we employed the morphological models described in Sec. \ref{sec:host_morph} as templates for image subtraction. The extreme obscuration along the line of sight led to non-detections at all UV/optical wavelengths for the transient, with the nIR counterpart to \grb\ only detected in $H$, $F160W$, and $K_s$ bands. The host galaxy was also found to be highly obscured, but our observations revealed significant detections in $z$, $H$, $F160W$, and $K_s$ bands. This dataset was combined with publicly available UV/optical/IR data from the literature for our subsequent analysis (see Table \ref{tab:observations_table} for the full afterglow photometry dataset and Table \ref{tab:host_observations_table} for all measured host photometry).

\begin{figure*}[htbp]
    \centering
    \includegraphics[width=\linewidth]{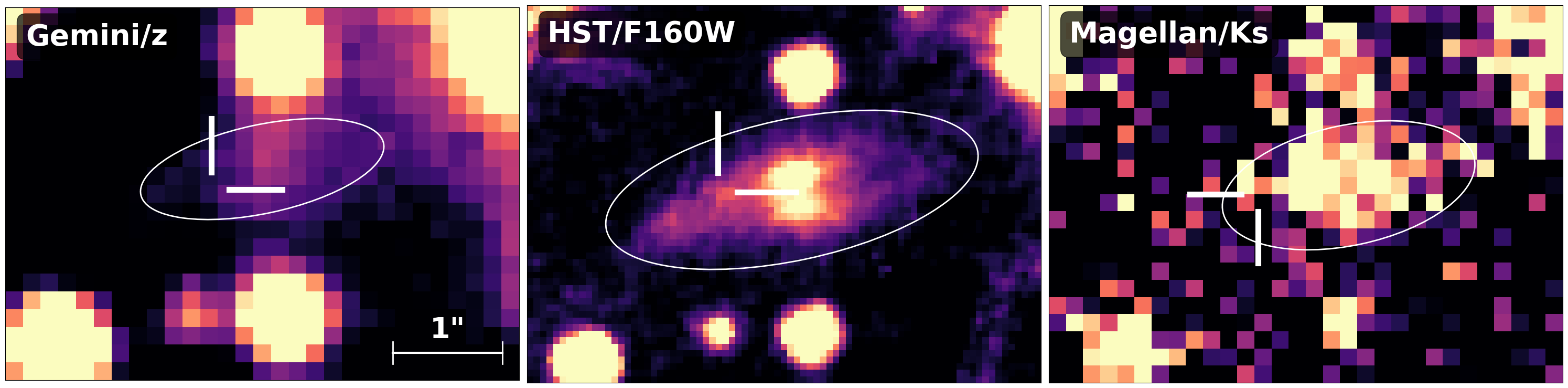}
    \caption{
    The host as detected by Gemini in $z$ band (left), \textit{HST} in the $F160W$ filter (similar to $H$-band; center) and Magellan in $K_s$ band (right). In all three images North is up and East is to the left. These are the host detections used for SED modeling (Sec. \ref{sec:sed_modeling}). In each image a crosshair marks the location of the afterglow.
    }
    \label{fig:hostfig}
\end{figure*}

\begin{table*}[htbp]
    \centering
        \caption{Photometry table of UVOIR observations of \grb. All magnitudes are provided in AB system and are not corrected for Galactic extinction. 
        }
    \label{tab:observations_table}
    \begin{tabular}{c|c|c|c|c}
    \hline
    \hline
         Datetime & Instrument & Filter & Transient Mag & Reference\\
\hline         
2025-07-03 02:06:10 & GOTO & $L$ & $>$ 20.35 & \cite{GOTO_GCN_EP250702A}\\
2025-07-03 05:12:58 & Blanco/DECam & $g$ & $>$ 23.56 & This work\\
2025-07-03 05:18:18 & Blanco/DECam & $r$ & $>$ 23.04 & This work\\
2025-07-03 05:22:05 & Blanco/DECam & $i$ & $>$ 22.26 & This work\\
2025-07-03 05:24:03 & Blanco/DECam & $z$ & $>$ 22.21 & This work\\
2025-07-03 05:26:10 & Blanco/DECam & $z$ & $>$ 22.08 & This work\\
2025-07-03 07:03:03 & VLT/HAWK-I & $H$ & 20.78 ± 0.05 & \cite{Levan_long_GRB}\\
2025-07-03 07:39:40 & VLT/HAWK-I & $K$ & 19.36 ± 0.02 & \cite{Levan_long_GRB}\\
2025-07-03 08:19:00 & COLIBRÍ/DDRAGO & $i$ & $>$ 22.0 & \cite{Becerra_long_GRB}\\
2025-07-03 10:54:08 & SVOM/VT & $R$ & $>$ 23.6 & \cite{SVOM_epoch_1}\\
2025-07-03 10:54:08 & SVOM/VT & $B$ & $>$ 23.8 & \cite{SVOM_epoch_1}\\
2025-07-03 17:26:10 & WFST & $r$ & $>$ 22.2 & \cite{Hua_long_GRB}\\
2025-07-03 19:36:00 & SYSU 80cm & $J$ & $>$ 17.5 & \cite{Li_long_GRB}\\
2025-07-03 21:52:47 & FTW/3KK & $r$ & $>$ 22.47 & This work\\ 
2025-07-03 21:52:47 & FTW/3KK & $i$ & $>$ 22.04 & This work\\ 
2025-07-03 21:52:47 & FTW/3KK & $J$ & $>$ 21.25 & This work\\ 
2025-07-03 22:50:00 & NOT/StanCAM & $i$ & $>$ 24.9 & \cite{Levan_long_GRB}\\
2025-07-03 23:45:36 & CAHA2.2/CAFOS & $z$ & $>$ 21.4 & \cite{Perez-Garcia_long_GRB}\\
2025-07-04 01:50:20 & Swift/UVOT & $b$ & $>$ 19.83 & \cite{Siegel_uvot_long_GRB}\\
2025-07-04 01:48:55 & Swift/UVOT & $u$ & $>$ 19.73 & \cite{Siegel_uvot_long_GRB}\\
2025-07-04 01:57:09 & Swift/UVOT & $v$ & $>$ 18.44 & \cite{Siegel_uvot_long_GRB}\\
2025-07-04 01:46:11 & Swift/UVOT & $uvw1$ & $>$ 19.93 & \cite{Siegel_uvot_long_GRB}\\
2025-07-04 01:51:45 & Swift/UVOT & $uvw2$ & $>$ 20.12 & \cite{Siegel_uvot_long_GRB}\\
2025-07-04 03:20:59 & VLT/HAWK-I & $K$ & 20.85 ± 0.03 & \cite{Levan_long_GRB}\\
2025-07-04 03:36:09 & VLT/HAWK-I & $H$ & 22.39 ± 0.06 & \cite{Levan_long_GRB}\\
2025-07-04 03:43:54 & Gemini/GMOS-S & $z$ & $>$23.95 & This work\\ 
2025-07-04 05:32:54 & Swift/UVOT & $uvm2$ & $>$ 19.33 & \cite{Siegel_uvot_long_GRB}\\
2025-07-04 22:40:29 & FTW/3KK & $r$ & $>$ 22.06 & This work\\ 
2025-07-04 22:40:29 & FTW/3KK & $i$ & $>$ 21.79 & This work\\ 
2025-07-04 22:40:43 & FTW/3KK & $K_s$ & 21.07 ±  0.16 & This work\\ 
2025-07-05 01:44:52 & GTC/HiPERCAM & $g$ & $>$ 23.6 & \cite{Levan_long_GRB}\\
2025-07-05 01:44:52 & GTC/HiPERCAM & $r$ & $>$ 23.0 & \cite{Levan_long_GRB}\\
2025-07-05 01:44:52 & GTC/HiPERCAM & $i$ & $>$ 22.6 & \cite{Levan_long_GRB}\\
2025-07-05 01:44:52 & GTC/HiPERCAM & $z$ & $>$ 22.3 & \cite{Levan_long_GRB}\\
2025-07-05 03:07:37 & VLT/HAWK-I & $K$ & 21.49 ± 0.05 & \cite{Levan_long_GRB}\\
2025-07-05 08:56:18 & SVOM/VT & $R$ & $>$ 23.6 & \cite{SVOM_epoch_2}\\
2025-07-05 08:56:18 & SVOM/VT & $B$ & $>$ 23.8 & \cite{SVOM_epoch_2}\\
2025-07-05 10:14:00 & Keck/MOSFIRE & $K_s$ & 21.96 ± 0.1 & This work\\ 
2025-07-05 10:14:00 & Keck/MOSFIRE & $H$ & $>$ 22.12 & This work\\ 
2025-07-05 10:14:00 & Keck/MOSFIRE & $J$ & $>$ 22.29 & This work\\ 
2025-07-06 02:16:22 & GTC/EMIR & $K_s$ & $>$ 21.5 & \cite{Levan_long_GRB}\\
2025-07-08 02:43:27 & GTC/EMIR & $K_s$ & $>$ 22.0 & \cite{Levan_long_GRB}\\
2025-07-08 06:50:09 & VLT/HAWK-I & $K$ & 22.53 ± 0.06 & \cite{Levan_long_GRB}\\
2025-07-10 19:26:50 & NEWFIRM & $H$ & $>$ 20.50 & This work\\ 
2025-07-11 21:33:30 & NEWFIRM & $H$ & $>$ 20.10 & This work\\
2025-07-11 21:50:51 & NEWFIRM & $K_s$ & $>$ 20.62 & This work\\ 
2025-07-15 01:40:27 & VLT/HAWK-I & $K$ & 22.82 ± 0.08 & \cite{Levan_long_GRB}\\

2025-07-15 02:39:23 & HST/WFC3 & $F160W$ & 26.52 ± 0.10 & This work\\
2025-07-20 09:55:45 & Gemini/GMOS-N & $z$ & $>$ 24.18 & This work\\ 
2025-08-12 06:00:00 & Magellan/FourStar & $K_s$ & $>$ 21.94 & This work\\
\hline
    \end{tabular}
\end{table*}

\begin{figure*}
    \centering
    \includegraphics[width=2.1\columnwidth]{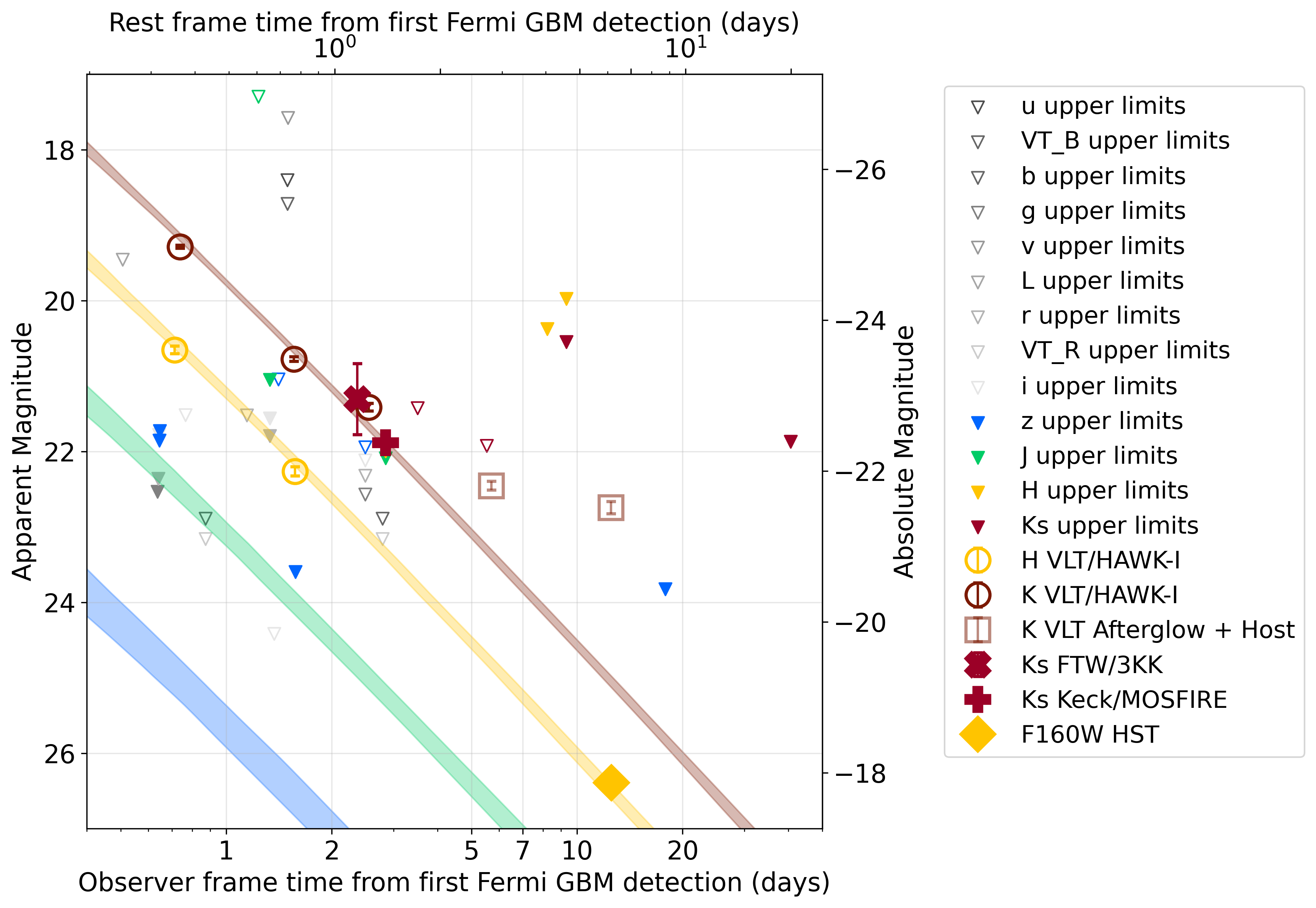}
    \caption{ 
    Light curve of the observed infrared counterpart of \grb, including UV and optical upper limits.
    $t=0$ is defined as the time from 2025-07-02 13:09:02 in the observer frame.
    Rest-frame time from the first \textit{Fermi} trigger and absolute magnitudes are calculated at $z=1.036$ \citep{Gompertz2025} without additional K-correction and are corrected for Milky-Way extinction using the dust maps by \cite{Schlafly2011}. Error bars represent $1\sigma$ confidence and upper limits represent $5\sigma$ confidence. UV/optical observations are colored in grayscale with lighter colors representing longer wavelengths; near-infrared observations are colored blue to red according to their wavelengths. Data presented for the first time in this work, together with our host-model-subtracted photometry of the \textit{HST} data first published in \cite{Levan_long_GRB}, are represented by solid markers, whereas data from the literature are represented by unfilled markers (see Table \ref{tab:observations_table}). Later VLT epochs with likely host contamination are denoted separately as squares and were excluded from modeling.} For $z$, $J$, $H$, $K$-bands, the 1$\sigma$ region in the posteriors for the best-fit forward shock model described in Section \ref{sec:jet} is shown as a shaded region.
    \label{fig:photometry}
\end{figure*}

\section{Analysis}
\label{sec:analysis}

\subsection{Jet Modeling}
\label{sec:jet}

We used the observed nIR detections obtained with Keck/MOSFIRE, FTW, VLT/HAWK-I, and \textit{HST} for model fitting (see Table \ref{tab:observations_table} and Figure \ref{fig:photometry}). We fit the $H$ and $K$-band data out to $3$ days after discovery. We neglected the last two VLT/HAWK-I $K$-band detections as they are likely contaminated by host galaxy flux \citep{Levan_long_GRB} and require late-time host galaxy templates to fully capture the transient flux.
We first fitted a powerlaw,
\begin{equation}
F(\nu,t) = A\nu^{\beta}(t - t_0)^\alpha,
\end{equation}
where $\nu$ is the effective frequency associated with each filter, $\beta$ is the spectral index, $\alpha$ is the temporal index, and $A$ is the normalization.
While the exact start time of the afterglow (the jet launching time) for the forward shock could have been coincident with the \textit{Fermi} B, D and E bursts \citep{Neights2025}, here we assumed the time of the first burst, GRB 250702D, as the initial time $t_0$.  From this joint temporal and spectral fit, we obtained a temporal index $\alpha=-1.83^{+0.05}_{-0.06}$ and spectral index $\beta=-4.95^{+0.32}_{-0.32}$. In particular, we find that the late-time \textit{HST} detection, after removing the majority of the host contamination (Section \ref{HST}) is consistent with the single powerlaw decay shown at earlier times. Overall, the near-infrared temporal index is consistent with the decay of the X-ray lightcurve \citep{OConnor2025}, but the observed spectral index is clearly inconsistent with the initial X-ray reports \citep[$\beta=-0.65\pm0.20$;][]{kenneagcn}.

Assuming the observed near-infrared emission arises from synchrotron radiation produced by the forward shock (the afterglow) of a relativistic jet \citep{Granot2002}, we can infer the properties of the blastwave and its surrounding environment. Using the standard afterglow closure relations \citep{Granot2002}, we can infer the slope $p$ of the electron's powerlaw energy distribution $N(\gamma)\propto \gamma^{-p}$. Assuming the near-infrared emission lies below the cooling frequency $\nu_\textrm{c}$ implies $p=1-2\beta=10.9\pm0.64$ whereas if the near-infrared is above the cooling frequency we have $p=-2\beta=9.90\pm0.64$. Either of these values are significantly inconsistent with both the theoretical and observed values of $p$ which lie closer to $p\sim2.3$ \citep{2001MNRAS.328..393A,2002APh....18..213E,2008ApJ...682L...5S,2010ApJ...716L.135C,2015ApJS..219....9W}. This demonstrates that the spectral index is incredibly steep and is strongly indicative of intrinsic dust extinction \cite[see also][]{Levan_long_GRB,OConnor2025} in addition to the large component in the Milky Way which was already corrected for \citep{Schlafly2011}. 

Therefore, we consider the closure relations utilizing the temporal decay index. This depends on the slope of the surrounding environment's density profile of $\rho(r)\propto r^{-k}$, where $k=0$ refers to a homogeneous medium and $k=2$ refers to a wind environment \citep{1998MNRAS.298...87D,2000ApJ...536..195C}. We start by considering the case of a pre-jet-break afterglow. Assuming a homogeneous medium, if the UVOIR data satisfy $\nu_m<\nu<\nu_c$, where $\nu_m$ is the injection frequency, this value of $\alpha$ implies an electron powerlaw index of $p = 3.44^{+0.07}_{-0.08}$. In a wind environment, this instead yields $p = 2.77^{+0.07}_{-0.08}$. If the UVOIR data lies above the cooling frequency, instead we have  $p = 3.11^{+0.07}_{-0.08}$, independent of the surrounding environment. While these values of $p$ are high, they are not inconsistent with the broad range of values found for GRB afterglows. 

If we instead consider a post-jet-break afterglow from a non-spreading jet \citep[e.g.,][]{Levan_long_GRB,OConnor2025}, for $\nu_m<\nu<\nu_c$, we find values of $p = 2.44^{+0.07}_{-0.08}$ and $p = 2.11^{+0.07}_{-0.08}$ for a homogeneous and wind environment, respectively. As it is unlikely that the near-infrared lies above the cooling frequency, we consider only this solution. As both of these values are typical of GRB afterglows, we cannot differentiate between these environments based on the simple temporal fit alone. Specifically, the inclusion of an intrinsic dust component is required to derive the reasonable range of intrinsic spectral slopes, and therefore aid in constraining $p$. 

Therefore, we used \texttt{VegasAfterglow} \citep{2025arXiv250710829W} to fit the observed UVOIR data to assess whether it is consistent with synchrotron emission arising from the forward shock of a top-hat jet \citep{Sari1998,Granot2002}.  We do this using \texttt{bilby} \citep{2019ApJS..241...27A} with nested-sampling using \texttt{dynesty} \citep{dynesty}.  We assumed a wind medium ($k=2$), as favored in other recent works \citep{Levan_long_GRB,OConnor2025}, and fit for $A_*$, the number density of the wind environment measured at a radius of $10^{18}$ cm. 
We assumed an on-axis viewing angle, $\theta_{\mathrm{obs}}=0$, because \grb's multi-wavelength light curve does not exhibit features suggestive of off-axis viewing—the X-ray light curve decays monotonically without the delayed rebrightening characteristic of off-axis jets.
We fit for the jet's core half-opening angle, $\theta_\textrm{c}$.
The free parameters in the fit are the isotropic-equivalent kinetic energy at the jet's core $E_\textrm{kin,iso}$, the initial bulk Lorentz factor at the jet's core $\Gamma_0$, density $A_*$, jet's half-opening angle $\theta_\textrm{c}$, slope of the electron's energy distribution $p$, and the electron ($\epsilon_e$) and magnetic field ($\epsilon_B$) energy fractions. We included an additional intrinsic dust component to our modeling using the Milky Way dust law reported in \citet{2007ApJ...663..320F}, which is parametrized by $A_{\textrm{V},\mathrm{host}}$ with fixed $R_\textrm{V}=3.1$. This dust component $A_{\textrm{V},\mathrm{host}}$ was uniformly sampled between 0 and 10 magnitudes.  We did not include UVOIR upper limits in the fit, and focused on the $H$ and $K_s$-band detections (Figure \ref{fig:photometry}. We supplemented the UVOIR dataset (Table \ref{tab:observations_table}) with additional X-ray and radio data \citep{OConnor2025,Levan_long_GRB}.

\begin{figure*}
    \centering
    \includegraphics[width=2.1\columnwidth]{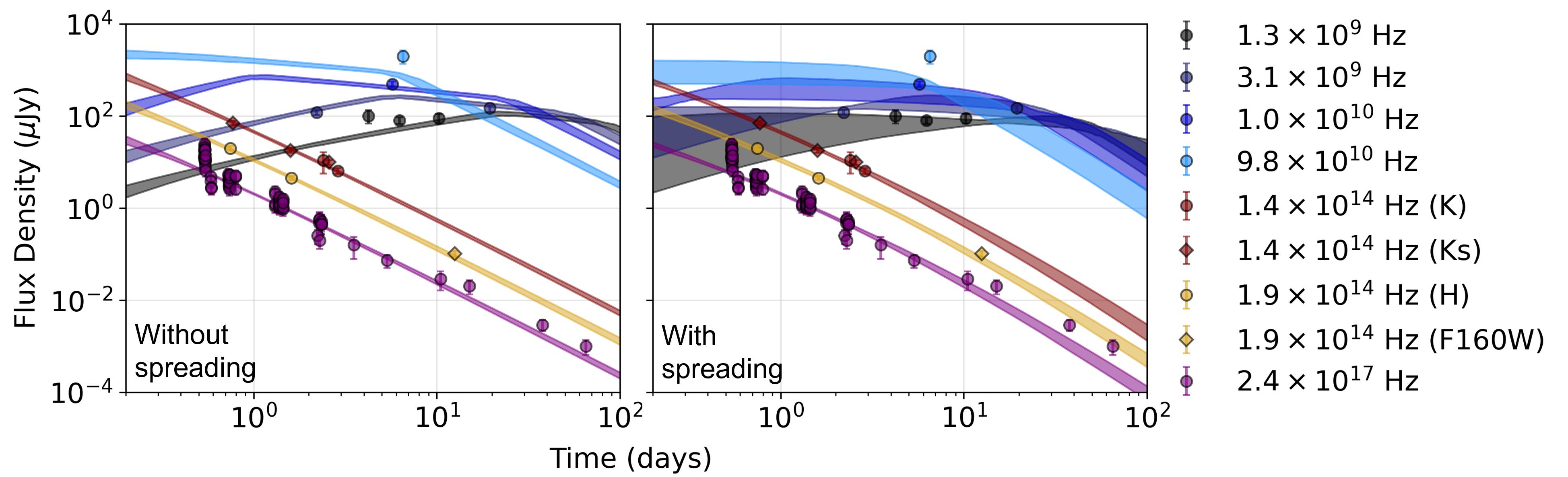}
    \caption{Radio, X-ray and infrared detections of the counterparts to \grb. The $1\sigma$ posteriors from \texttt{VegasAfterglow} fit to these data are shown as shaded regions. The left-hand panel shows the fitting results without jet spreading enabled and the right-hand panel shows the results with jet spreading enabled.}
    \label{fig:multiwav_model}
\end{figure*}

We performed three \texttt{VegasAfterglow} fits to the full multi-wavelength (X-ray to near-infrared to radio) lightcurve using \texttt{bilby} \citep{2019ApJS..241...27A} with nested-sampling using \texttt{dynesty} \citep{dynesty}. The upper limits were not included because they are unconstraining for the fit: in the UV/optical bands this is due to the large amount of dust along the line of sight, and so are the nIR upper limits because of lack of depth (see Fig\,\ref{fig:photometry}). The first fit was ran with the exclusion of lateral jet spreading effects and the second fit included jet spreading \citep[see][for details]{2025arXiv250710829W}. A third fit for a non-spreading jet was performed with the added requirement that $\Gamma_0\theta_\textrm{c}>1$ such that the spherical jet approximation holds during the deceleration phase \citep[e.g.,][]{Rhoads97,Rhoads1999}. The fit results are shown in the corner plots in Figures \ref{fig:jet_corner_plot-nospread}, \ref{fig:jet_corner_plot-spread}, and \ref{fig:jet_corner_plot-constraint}. The model comparison to the observed multi-wavelength lightcurve is shown in Figure \ref{fig:multiwav_model} for both the non-spreading (without the requirement of $\Gamma_0\theta_\textrm{c}>1$) and spreading models. We find a Bayes factor (BF) strongly in favor of the model without jet spreading enabled of $\log_{10}\mathrm{BF}=26.8$. We discuss these solutions in \S \ref{sec:jetenergy}.  

There are significant departures from our models in radio wavelengths.  However, at these observing frequencies, scintillation can induce high variability  \citep[see][and the references therein]{2020MNRAS.494.2449D}.  We therefore cannot conclude that these are intrinsic outliers.  Additionally, the early 1\,keV light curve departs significantly from the fitted models which is interpreted to be flaring activity in \citet{OConnor2025}.

\begin{figure*}
    \centering
    \includegraphics[width=2.05\columnwidth]{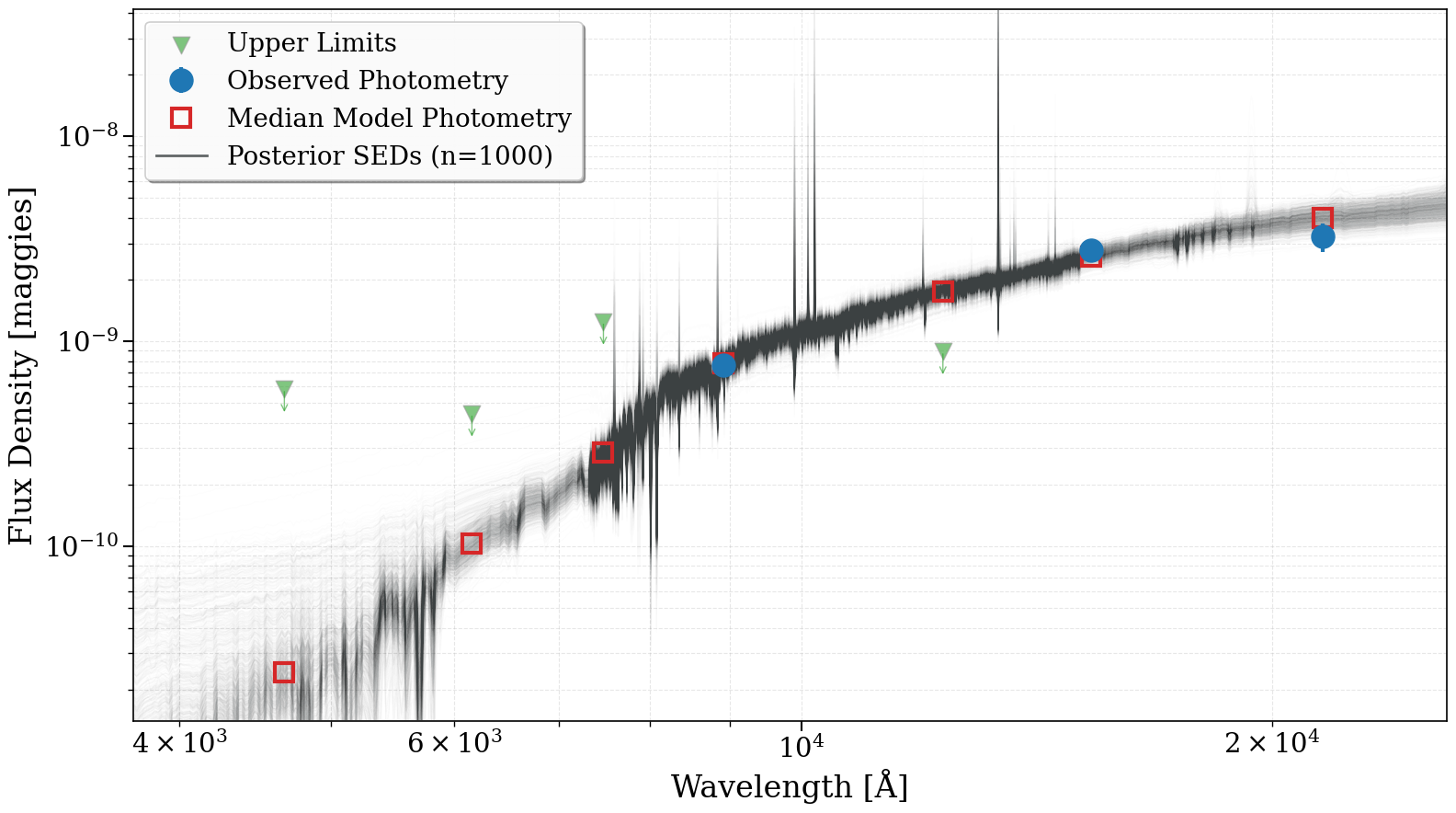}
    \includegraphics[width=2.05\columnwidth]{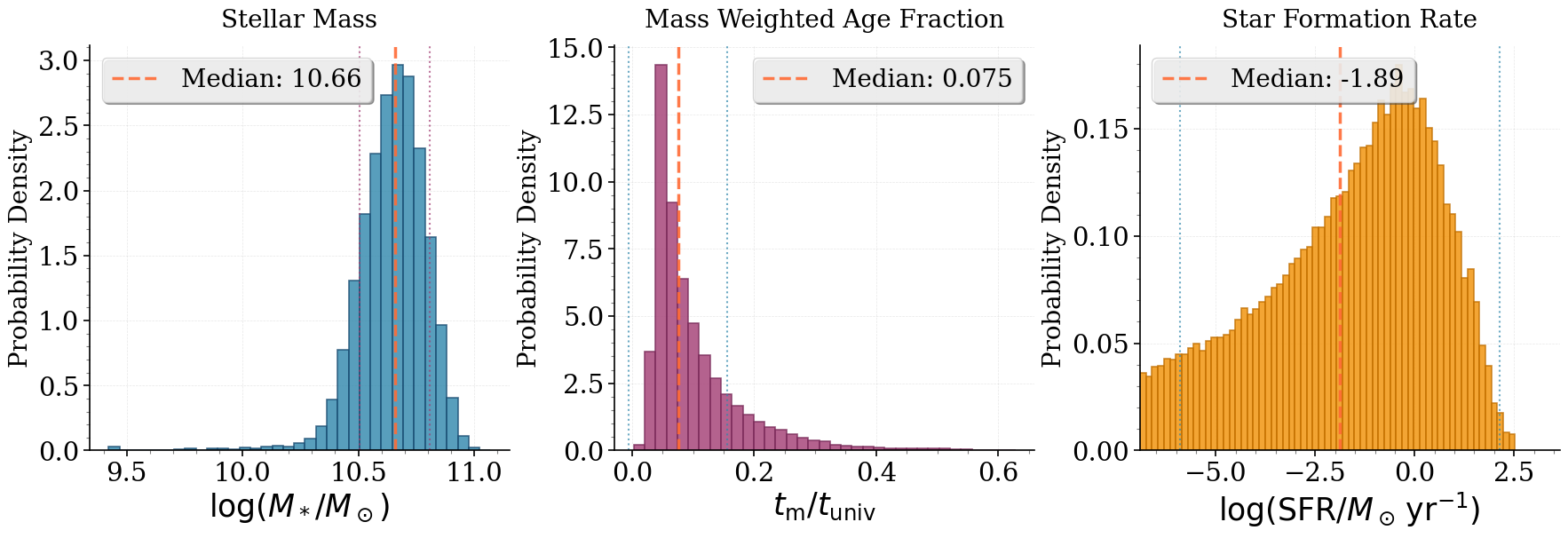}
    \caption{
    \textit{Top panel:} Observed host photometry ($z$, $F160W$, $K_s$ with $g$ and $r$ upper limits) overlaid with 1000 randomly drawn galaxy SEDs from the \texttt{prospector} posterior distributions (Figure \ref{fig:prospector_corner_plot}). The data are corrected for Milky Way extinction prior to modeling. \textit{Bottom left:} Posterior distribution of stellar mass from \texttt{prospector} modeling. \textit{Bottom center:} Posterior distribution of mass-weighted age from \texttt{prospector} modeling. Here we show it in units of $5.71$ Gyr (the age of the Universe at $z=1.036$). \textit{Bottom right:} Posterior distribution for star formation rate from \texttt{prospector} modeling.
    }
    \label{fig:prospector}
\end{figure*}

\subsection{Host Galaxy SED Modeling}
\label{sec:sed_modeling}

The host galaxy of \grb\ is extremely red and was only detected in the $zHK$ filters (Figure \ref{fig:hostfig}). 
Due to the diffuse nature of the host galaxy (Figure \ref{fig:host_morphology}) the standard point source upper limits derived in other filters are misleading, thus we exclude the $i$ and $J$ band upper limits from the model fitting. We choose to include $g$ and $r$ band upper limits (Table \ref{tab:observations_table}) as they are half a mag deeper and thus more likely to still be constraining on a highly extincted SED.
We modeled the spectral energy distribution (SED, top panel of Figure \ref{fig:prospector}) using the $zHK$ host detection (Table \ref{tab:host_observations_table}) with the flexible stellar population (FSPS) code \citep{Conroy2009}. We utilized  \texttt{prospector} \citep{Leja2017,Johnson2019,Johnson2021} in accordance
with the methods previously outlined in \citet{OConnor2022}. 
The model fit was performed with \texttt{emcee} \citep{emcee}. 
We adopted a \citet{Chabrier2003} initial mass function, a \citet{Calzetti2000} dust law, and a delayed-$\tau$ star formation history.  
We fit for the total mass $M_\textrm{gal}$, the age of the host galaxy $t_{\rm age}$, e-folding timescale $\tau$, metallicity $Z$, and intrinsic dust $A_{\textrm{V},\mathrm{host}}$. We adopt uniform bounded priors with ranges of 0.01--5.71 Gyr for $t_{\rm age}$, 0.1--10.0 Gyr for $\tau$, $-1.8$ to $+1.2$ dex (relative to solar) for log($Z/Z_\odot$), and 0--25 mag for dust optical depth. 
We then convert these into posterior distributions for total stellar mass, mass-weighted stellar age, 
and star formation rate (SFR, Fig \ref{fig:prospector}, bottom panels). 
Due to the sparsity of our observations we poorly constrained all parameters except for total stellar mass and dust, for which we find a massive ($M_*=10^{10.66\pm0.15}\ M_\odot$) and dusty ($A_{\textrm{V},\mathrm{host}}=0.88^{+0.47}_{-0.45}$ mag) host. A corner plot showing the posterior distributions is presented in Figure \ref{fig:prospector_corner_plot}. We note that without the addition of redder (rest frame) wavelengths in the fit the inferred mass may be lower than the true stellar mass of the galaxy.

\subsection{Morphological Analysis of the Host Galaxy}
\label{sec:host_morph}

\textit{HST} observations of the host galaxy of \grb\ revealed a peculiar, disturbed morphology. We performed a full morphological analysis using both parametric and non-parametric fitting techniques to investigate the underlying features in the galaxy's light. 

\begin{figure*}
    \centering
    \includegraphics[width=\linewidth]{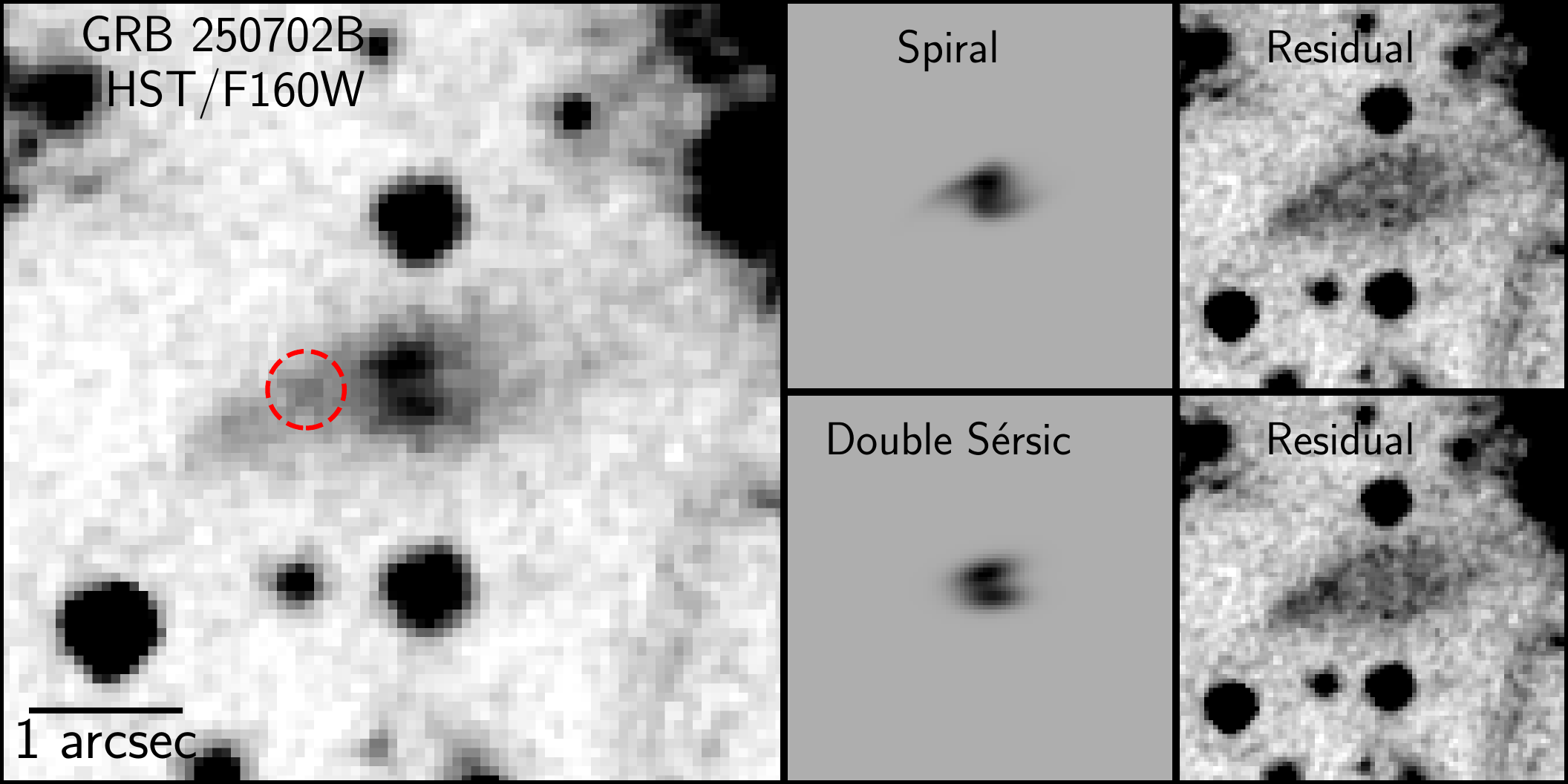}  
    \caption{
     \textit{Left panel:} \textit{Hubble Space Telescope} image of the host galaxy of \grb\ in the $F160W$ filter. The location of the transient is marked by a red circle. \textit{Top Middle panel:} \texttt{GALFIT} model of best-fitting single Sérsic light profile and powerlaw spiral arms. \textit{Top Right panel:} The residual image of \texttt{GALFIT} after subtracting the spiral plus Sérsic model. \textit{Bottom Middle panel:} \texttt{GALFIT} model of best-fitting double Sérsic light profiles. \textit{Bottom Right panel:} The residual image of \texttt{GALFIT} after subtracting the Double Sérsic model. In all images North is up and East is to the left
    }
    \label{fig:host_morphology}
\end{figure*}

We began with a parametric analysis using \texttt{GALFIT} \citep{Peng2002}. We produced a bad-pixel segmentation map using \texttt{Photutils} \citep{Bradley_2024} to mask out nearby objects that are unrelated to the host galaxy, and a PSF using the \texttt{Spike} software \citep{Polzin2025}. As the morphology of the galaxy clearly shows two separate components, we performed a fit using two Sérsic profiles (Figure \ref{fig:host_morphology}). We derive a half-light radius of the upper (North) Sérsic of $r_{50} = 0.31'' \pm 0.08''$ and the lower (South) Sérsic of $r_{50} = 0.30'' \pm 0.11''$. The Sérsic indices are $0.56 \pm 0.41$ and $0.49 \pm 0.44$, respectively, indicative of disky galaxies.

An additional fit was performed to investigate any possible underlying spiral structure. We applied two Sérsic profiles stacked onto the same coordinates, with applied azimuthal profile functions on the top profile. The primary Sérisic captures the central light, while the secondary Sérsic fits to the spiral structure. The ellipsoid shape of the galaxy light was perturbed using three functions: Fourier modes, bending modes, and a power-law spiral. The fit is shown in Figure \ref{fig:host_morphology}. For this model, the derived half-light radius for the central Sérisic is $r_{50} = 0.18'' \pm 0.19''$ and $r_{50} = 0.45'' \pm 1.18'' $ for the second Sérsic. The Sérsic indicies are $0.31 \pm 0.95$ and $0.41 \pm 0.64$ respectively.

These two parametric fits yield the same goodness-of-fit and reveal similar residuals to the underlying structure. As such we cannot distinguish between these two models, and suggest further imaging with \textit{JWST} is required. In both fits, we find that after subtracting the \texttt{GALFIT} models there is a clear diffuse, low surface brightness emission that extends in the East-West direction above and below the galaxy's center for $\sim$\,$1.75\arcsec$.
This may be due the possible merger of the two separate Sérsic profiles, and does offer some support to that scenario, though it is still unclear whether the galaxy is two separated, possibly merging, Sérsic profiles, or a single galaxy separated by a dust lane \citep{Levan_long_GRB}. Notably, a residual consistent with a point source is visible in both cases at the transient location.

\begin{figure*}[htbp]
    \centering
    \includegraphics[width=\linewidth]{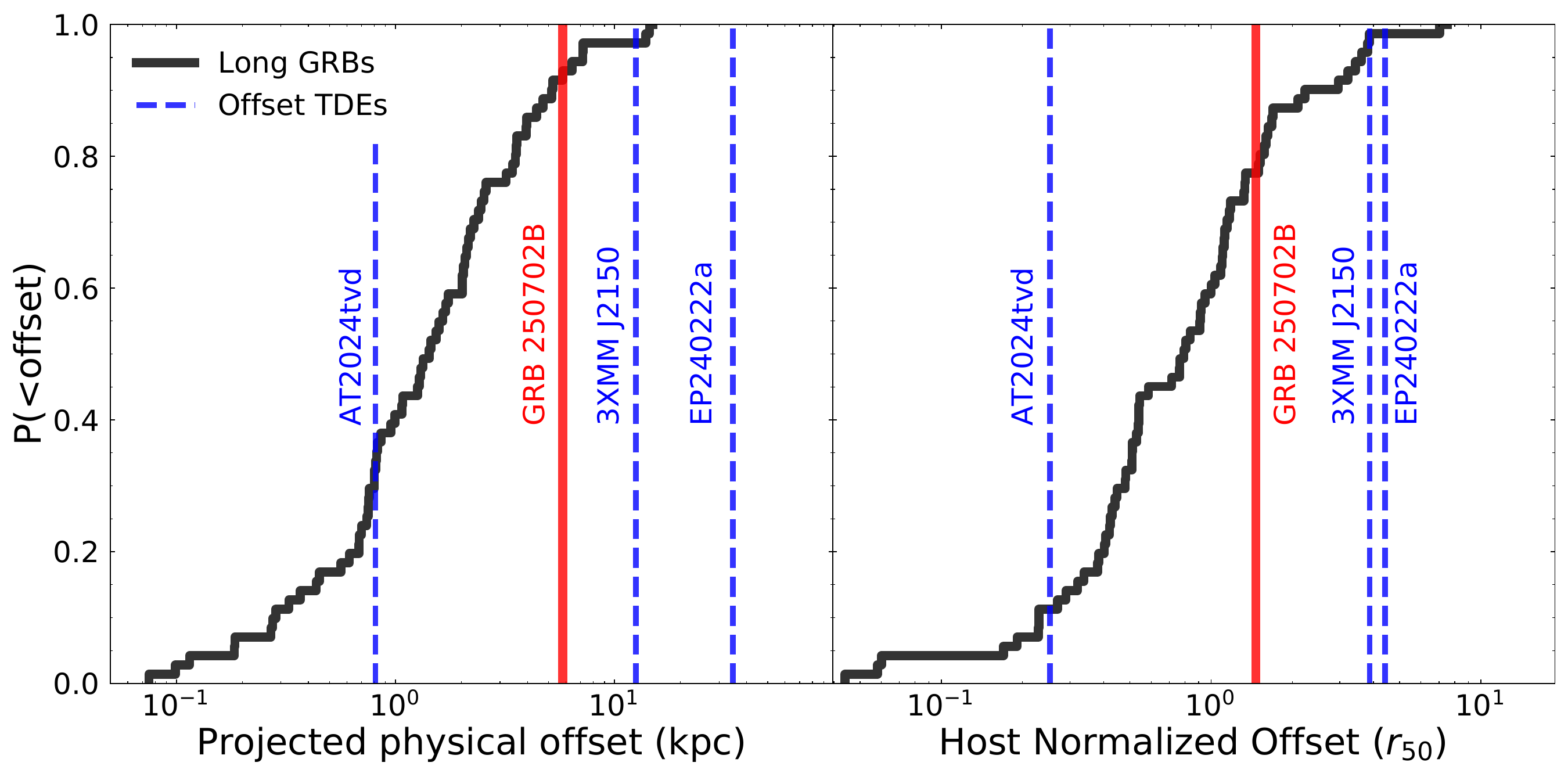}
    \caption{\textbf{Left:} The projected galactocentric offset of \grb\ compared to  offset TDEs \citep{Yao2025ApJ} and the cumulative distribution for long GRBs \citep{Blanchard2016}. \textbf{Right:} The same but for the host normalized offset ($r/r_{50}$). The offset TDE candidates 3XMM J2150 \citep{Lin2018, Lin2020}, EP240222a \citep{Jin2025}, and AT2024tvd \citep{Yao2025ApJ} are believed to be associated with an IMBH.
    }
    \label{fig:offset}
\end{figure*}
\begin{figure*}[ht!]
    \centering
    \includegraphics[width=\linewidth]{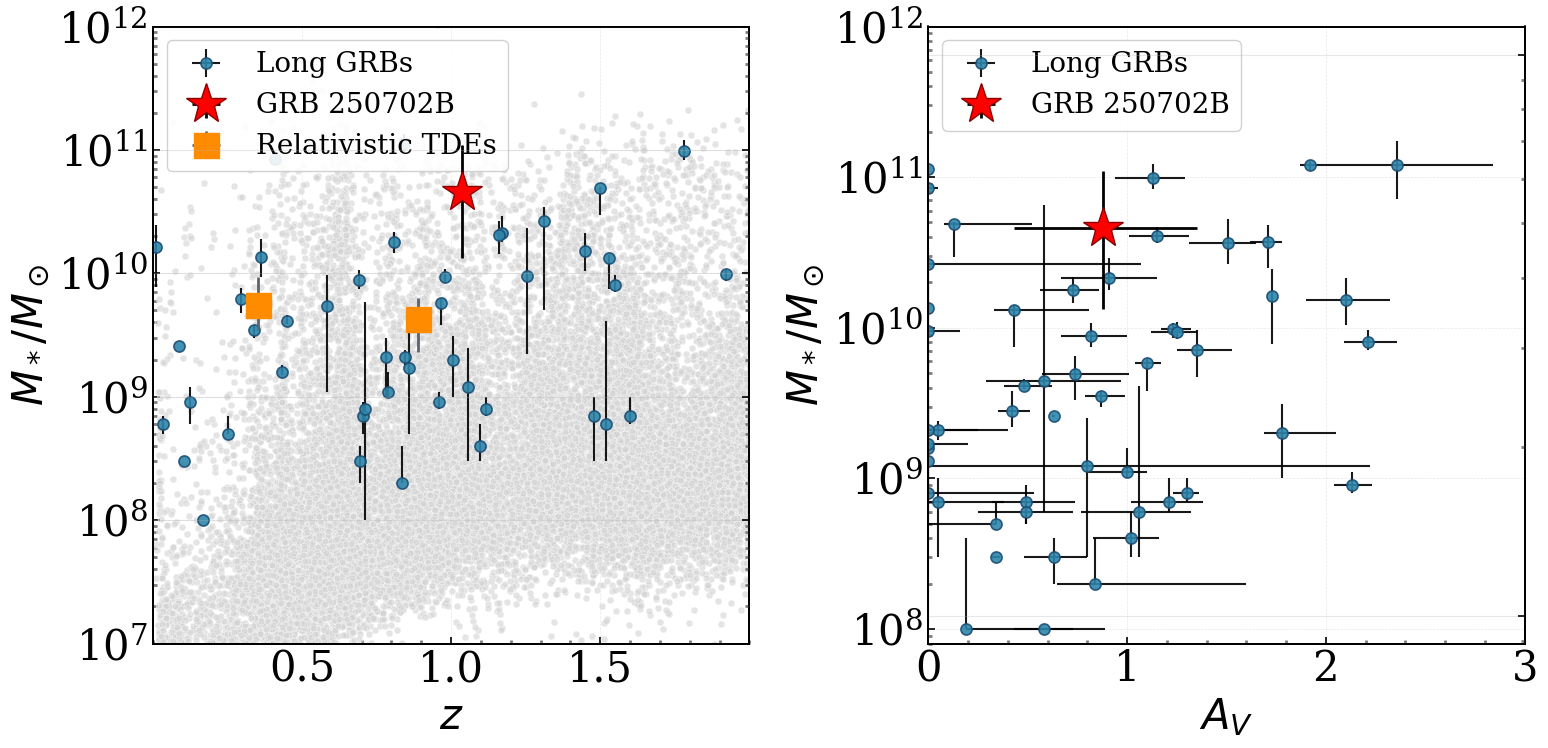}
    \caption{\textbf{Left:} Stellar mass versus redshift for long GRB host galaxies \citep[blue circles;][]{Savaglio2009,Perley2013} and field galaxies from the CANDELS survey \citep[gray circles;][]{Galametz2013}. Also shown are the two relativistic TDEs (orange squares): Sw J1644+57 \citep{Yoon2015,Levan2016ApJ} and Swift J1112.2-8238 \citep{Brown2015,Brown2017}. \grb\ is shown as a red star. \textbf{Right:} Stellar mass versus the average host extinction inferred from galaxy SED modeling \citep{Perley2013}.
    }
    \label{fig:host_mass_and_dust_comparison}
\end{figure*}

To further investigate the host's nature, we performed a non-parametric analysis using the \texttt{Morfometryka} software \citep{Ferrari2015}. This makes use of the three-dimensional ``CAS'' volume, representing the the concentration (C), asymmetry (A), and clumpiness (S) of a galaxy's light profile, to classify a given galaxy's morphology \citep[e.g.,][]{Conselice2000,Conselice2003}. 
For this galaxy, we derive a concentration of C$=2.56_{-0.06}^{+0.05}$ and an asymmetry of $A=0.90_{-0.05}^{+0.05}$. The concentration is determined by C$=5\log_{10}(r_{80}/r_{20})$ and the asymmetry is determined by rotating the image about its center, subtracting the rotated image from the initial image and then measuring the degree of residuals left behind. The classification regions defined by \cite{Bershady_2000} determined that a high asymmetry of $A > 0.35$ indicates a galaxy undergoing a major merger. 
With an unusually large value of A, \grb\ was found to be highly asymmetrical, suggesting a significant merger is potentially taking place.

\section{Discussion}
\label{sec:discussion}

\subsection{Constraints on rebrightening}

Continued nIR observations extending to 40.67 (19.98) days in the observer (rest) frame post-\textit{Fermi} detection show no evidence of rebrightening (Figure \ref{fig:photometry}).
Our reported \textit{HST} F160W photometry is consistent with a continued power-law decay out to 12.53 (6.15) days in the observer (rest) frame (Figure \ref{fig:photometry}). This is in contrast to the non-power-law decay in the combined $H$ and \textit{F160W} lightcurve reported by \cite{Levan_long_GRB}, where no attempt was made to remove contaminating host flux from the \textit{HST} measurement. Similarly, the last two $K$-band VLT epochs in Figure \ref{fig:photometry} (denoted by squares) appear nonlinear in log-log scale but are similarly contaminated by host flux (see \cite{Levan_long_GRB} for a discussion about removing host flux from the VLT photometry). These points are not included in our jet modeling (\S \ref{sec:jet}). 
The final observation that we acquired, using the Magellan telescope, allows us to place only limited constraints on the rebrightening of the nIR transient. We are able to rule out rebrightening below an absolute magnitude of $K_s=-22.32$ AB mag at $\Delta t \approx 20$ days in the rest frame.
This is insufficiently deep to rule out rebrightening akin to either the well-studied optical jetted TDE AT2022cmc \citep{Andreoni2022, Pasham2023} or the prototypical GRB supernova SN1998bw \citep{Galma1998}.
AT2022cmc was observed in $K_s$ at rest frame $\Delta t=$ $3.02$, $9.02$ and $13.57$ days after the first ZTF detections with observed absolute magnitudes of $K_s=-24.01$, $K_s=-21.88$, and $K_s=-21.54$ AB mag respectively. Given \grb’s slightly lower redshift \citep[AT2022cmc occurred at $z=1.19325$;][]{Andreoni2022} and assuming the SED of AT2022cmc’s slow component takes roughly the form of a 30,000 K blackbody around a rest wavelength of 1,000 \AA, we would expect these values to be $\sim+0.3$ mags dimmer, making our late time observations $\sim 1.1$ mag too shallow to constrain this potentially slowly fading component of the UVOIR lightcurve. 
SN1998bw ($z=0.0085$) peaked in the $I$ band (the reddest wavelength for which photometry is available) at an absolute magnitude of $-18.82$ AB mag \citep[converted from Vega according to][]{Blanton2007} at $\Delta t\approx19$ days after the associated GRB.
Accounting for redshift, but not extinction or K-correction, this is roughly equivalent to an $H$ band observation of $H\approx 25.5$ AB mag for an object at $z=1.036$. 
Based on IR spectra taken within a few days of the $I$ band peak \citep{Patat2001} we infer that a constraining $K_s$ observation for a 1998bw-like supernova at the location of \grb\ would require observations deeper than $K_s\approx25.5$ AB mag.

\subsection{Dust extinction from jet modeling}

Our observations and analysis of the nIR transient further support its relativistic nature (\S \ref{sec:jet} and \ref{sec:jetenergy}), and therefore its association with the high-energy detections of \grb\  \citep{Neights2025, Levan_long_GRB, OConnor2025}. The bright peak absolute magnitude, rapid fading, and color evolution not only are well fit by relativistic jet models (Section\,\ref{sec:jet}), but are inconsistent with other common classes of extragalactic transients in the optical/nIR \citep[see peak magnitude vs. evolution time-scale plots for optical transients in, e.g.,][]{Kasliwal2011PhDT, Ho2022}.   

The source is very reddened due to high extinction, as also discussed by \cite{Levan_long_GRB,OConnor2025,Gompertz2025}. After correcting for Galactic extinction, we find that both the host galaxy and the jet models need to account for a significant amount of dust. The estimate for the extinction that we infer from the jet modeling ($A_{\textrm{V},\mathrm{GRB}} = 5.43^{+0.53}_{-0.47}$ mag) is significantly larger than the value we obtain from host galaxy SED fitting ($A_{\textrm{V},\mathrm{host}} = 0.88^{+0.47}_{-0.45}$ mag). This discrepancy suggests that the GRB was produced in a dense, dusty circumstellar environment, or a thick lane of dust is present in the host galaxy on the line of sight, between the observer and the transient \citep{Levan_long_GRB}. The line of sight extinction is comparable to the dustiest of those observed in a set of eight optically dark GRBs discussed in \cite[$A_\textrm{V} \gtrsim 3$ mag;][]{Kruhler2011_dustiest_GRBs} and to the heavily obscured relativisitc TDE Sw J1644+57 \citep[$A_\textrm{V} \approx4.5$ mag;][]{Burrows2011Nat}, see also \citet{OConnor2025}.

\subsection{Opening angle and collimation-corrected energetics}
\label{sec:jetenergy}

As discussed in \S \ref{sec:jet}, there exist two quantitatively 
different solutions for the afterglow model (see Figure \ref{fig:multiwav_model}) depending on the inclusion or exclusion of lateral jet spreading effects. 

The non-spreading jet fit (Figure \ref{fig:jet_corner_plot-nospread}) requires a narrow jet such that the lightcurve is already in a post-jet-break phase prior to the start of the earliest observations at $\sim$\,$0.5$ days from discovery. This provides a better match to the observed temporal slope and the required value of $p$ as noted above. This post-jet-break solution was also found by \citet{Levan_long_GRB,OConnor2025,Gompertz2025}. However, when including lateral jet spreading, the post-jet-break decay of $\approx$\,$t^{-p}$ \citep{Rhoads1999,SariPiranHalpern1999} is too steep to describe the observed temporal decline of $\alpha$\,$\approx$\,$-1.8$ found for both the optical and X-ray data. Therefore, the solution identified in the presence of lateral jet spreading (Figure \ref{fig:jet_corner_plot-spread}) requires that the jet-break occur either during or after our latest data point. The pre-jet-break temporal slope of the model is therefore slightly shallower ($\alpha$\,$\approx$\,$-1.45$ for $p$\,$\approx$\,$2.3$) than the observed steep X-ray and near-infrared decay $\alpha$\,$\approx$\,$-1.8$. Instead of identifying a steeper value of $p$ and placing the jet-break at $>65$ days, the fit using a spreading jet places the jet break earlier (at $\sim$\,$1-3$ days) and underproduces the late-time \textit{Chandra} X-ray detections \citep{OConnor2025}. In this case, while the jet is not extremely wide (as would be required for a $>65$ day jet-break; see \citealt{OConnor2023}), it is significantly wider compared to the non-spreading jet solution and therefore also requires a higher collimation-corrected energy. That said, the inferred opening angles and collimation-corrected energies (Figures \ref{fig:jet_corner_plot-nospread} and \ref{fig:jet_corner_plot-spread}) are consistent with the range observed from classical long GRBs \citep{Cenko2010,Wang2018,OConnor2023}. 

While at present the non-spreading post-jet-break solution provides a better fit to the data, there is no clear break observed in the lightcurve. As such, both solutions (wide jet versus narrow jet) for the afterglow are potentially valid (\S \ref{sec:jet}). While our afterglow modeling with a spreading jet (Figures \ref{fig:multiwav_model} and \ref{fig:jet_corner_plot-spread}) does not identify the late jet-break solution, an additional fit requiring a late jet-break ($>65$ days) was successfully able to model the lightcurve. This fit requires a steeper value of $p\approx2.6$ (with higher $A_\textrm{V}\approx8.5$ mag), but underpredicts the observed $H$-band flux, though a different dust law may account for this. Qualitatively the fit is viable and as the correct afterglow solution (wide jet versus narrow jet) has significant consequences for the progenitor's requirements, additional late-time X-ray observations and continued radio monitoring can constrain the presence of the onset of any jet-break in the lightcurve and possibly differentiate between these two possible solutions. 

\subsection{Constraints on the jet's Lorentz factor}

An additional critical constraint on the progenitor of GRB 250702B comes from the inferred bulk Lorentz factor $\Gamma_0$, which is found to be highly relativistic. Observing the end of the jet's deceleration phase is required to determine the exact Lorentz factor \citep{Sari1999,Molinari2007,Ghisellini2010}, but an upper limit to the deceleration time can also provide useful constraints \citep{OConnor2020,OConnor2024}. As we do not detect the deceleration of the afterglow (and therefore have an upper limit to the time of jet deceleration), the results of our fit can be considered a lower limit to the true Lorentz factor, as lower Lorentz factors will lead to later deceleration. 
The posterior distribution of \grb's initial Lorentz factor ($\log\Gamma_0=1.25^{+0.16}_{-0.12}$) in Figure \ref{fig:jet_corner_plot-nospread} favors ultrarelativistic blastwave with Lorentz factor $\gtrsim$\,$10$. 
Enforcing the condition $\Gamma_0\theta_\textrm{c}>1$ to prevent the jet from spreading laterally during the coasting phase requires a significantly larger Lorentz factor ($\log\Gamma_0=2.61^{+0.27}_{-0.41}$; see Figure \ref{fig:jet_corner_plot-constraint}).
We note that the detection of high energy $>1$ MeV photons by \textit{Fermi} sets an independent requirement of $\Gamma_0 \gtrsim 50-80$ \citep{Neights2025}. 
Measurements of $\Gamma_0$ of observed from relativistic jetted TDEs AT2022cmc and Swift J1644+57 lie around $\sim 8-20$ \citep{Zauderer2011,metzger12_j1644_afterglow,Andreoni2022, Rhodes2023}, which depart from the observed population of GRBs which generally have far larger Lorentz factors \citep[e.g., $\sim$ hundreds;][]{Liang2010,Ghirlanda2012,Ghirlanda2018}. 
As we cannot rule out an earlier deceleration of the jet, the measurements can be treated as a lower bound to the Lorentz factor. Higher Lorentz factors clearly disfavor a TDE-like progenitor scenario and are more consistent with a GRB-like progenitor (\S \ref{sec:prog}).

\subsection{Host galaxy properties and population comparisons}

In Figure \ref{fig:offset} we compare both the projected physical offset (\textit{left panel}) and host normalized offset (\textit{right panel}) of \grb\ with the population of classical long duration gamma-ray bursts \citep{Blanchard2016}. The location of \grb\ is displayed as a red vertical line and lies in the 90th percentile of long-GRB physical offsets and 80th percentile of host-normalized offsets. We find that the observed galactocentric offset of \grb\ also lies within the range of values observed for the population of candidate offset TDEs possibly arising from an intermediate mass black hole (IMBH) \citep{Lin2018, Lin2020,  Jin2025, Yao2025ApJ}. 
The offset alone does not distinguish between progenitor scenarios and is consistent with either a massive star or an offset IMBH progenitor.

Turning to the host parameters from SED modeling (Sec \ref{sec:sed_modeling}) we compare the inferred stellar mass as a function of both redshift and extinction to long GRB hosts. In the left panel of Figure \ref{fig:host_mass_and_dust_comparison} we compare \grb\ to both long GRB hosts \citep{Savaglio2009, Perley2013}, two relativistic TDEs with well-studied host galaxies \citep{Levan2016ApJ, Brown2015}, and field galaxies from the CANDELS survey \citep{Galametz2013}. We find that while the host of \grb\ is inferred to be massive compared to most field galaxies at a similar redshift and to both relativistic TDE hosts, its inferred stellar mass is consistent with that of the most massive long GRB hosts. 
In the right panel of Figure \ref{fig:host_mass_and_dust_comparison} we find that the inferred stellar mass and extinction are consistent with the most massive moderately obscured long GRB hosts \citep{Savaglio2009, Perley2013} and to the most massive of the eight non-optically selected ``dark" GRBs discussed in \citep{Kruhler2011_dustiest_GRBs}.
The inferred total stellar mass, $\log(M_*/M_\odot)=10.66\pm0.15$, is similar to the $\log(M_*/M_\odot)\approx10.9$ host mass of all three known off-nuclear TDEs \citep{Lin2018, Lin2020, Jin2025, Yao2025ApJ}. We caution that these comparisons are limited by the small populations of both optically dark GRBs and relativistic TDEs for which there are measured host properties.

In Section \ref{sec:host_morph} we derived the host's concentration (C$=2.56_{-0.06}^{+0.05}$) and asymmetry ($A=0.90_{-0.05}^{+0.05}$) and determined that the galaxy exhibits morphological features consistent with a major merger.
Using the CAS parameter space, we compared the host of \grb\ with the long GRB sample from \cite{Lyman2017}, which only yields hosts with $A < 0.4$.
The morphology of \grb's host is a significant outlier compared to the limited sample of morphologically characterized GRB hosts.
The morphology of \grb's host also differs significantly from the host of the relativistic TDE Sw J1644+57 \citep[A$\approx0.1$;][]{Levan2016ApJ}. 
However, the relativistic TDE Swift J1112.2-8238 occurred in a highly asymmetric and possibly merging galaxy \citep{Brown2017_J1112}. 
If confirmed, this merger scenario would be consistent with a TDE origin, as galaxy mergers are known to both enhance TDE rates \citep{Pfister2021} and produce populations of wandering massive black holes \citep{Ricarte2021, Tremmel2018}.

\subsection{Summary of the possible progenitors}
\label{sec:prog}

The gamma-ray duration of \grb\ suggest a similar origin to previously studied ultralong GRBs. While the core-collapse of a massive star could explain a $\sim 25$\,ks \citep{Neights2025} prompt emission, the observed long-lived, multi-day X-ray emission is incompatible with the standard collapsar models \citep{OConnor2025}. 
The off-center location of \grb\ is naturally consistent with stellar mass or intermediate mass black hole engines from several proposed scenarios: collapsars, compact object--helium star mergers \citep{Neights2025}, micro-TDEs \citep{Beniamini2025} or IMBH-TDEs \citep{Granot2025,Levan_long_GRB}. 
An association with a supernova would be expected in collapsars and some versions of micro-TDE origins involving a natal-kick-induced TDE after stellar collapse in one component of a field binary star system. 
An IMBH-TDE would also potentially require a supernova association. In this scenario, in the classic `frozen-in' approximation \citep{lacy82,Rees1988} energy dissipation takes place primarily at the tidal radius\footnote{If, instead, energy dissipation is dominated at the peri-center (for comparison of this with the `frozen-in' assumption, see \citealt{2019MNRAS.485L.146S}), and the penetration factor $\beta$ is rather large $\beta\gtrsim 5$, then the fallback time shortens significantly (as $\beta^{-3}$, see \citealt{Beniamini2025}) and disruption of a main sequence star by an IMBH becomes possible (albeit geometrically fine-tuned).} and the duration of fallback from the most inner-bound orbit is $t_{\rm fb}\sim 10^6(M_{\rm BH}/10^4M_{\odot})^{1/2}R_*^{3/2}M_*^{-1}\mbox{ s}$. As a result, matching the duration with that of GRB 250702B requires a smaller physical size of the disrupted star which can be realized if the star is a white dwarf. In this situation, the compression of a white dwarf during the TDE process is expected to be sufficient to trigger a thermonuclear supernova \citep{Rosswog2009}. 
Helium star mergers are also expected to have associated supernovae, though significantly dimmer than those associated with collapsars \citep{Neights2025}. 
Finally, the absence of a supernova would support a dynamically-induced micro-TDE. 
At the time of writing, the large redshift and significant dust extinction in \grb\ preclude a strong conclusion regarding potential supernova association.

\section{Conclusion}
\label{sec:conclusion}

We presented deep optical and nIR observations of the high energy transient \grb\ -- the longest duration GRB ever observed -- whose extreme multi-wavelength properties cannot be placed firmly among known transient classes
\citep{Levan_long_GRB, OConnor2025, Neights2025, Gompertz2025}.
Deep observations from a suite of optical/nIR facilities — including Blanco, Gemini, Wendelstein, Keck, and Magellan telescopes — combined with additional photometry from \cite{Levan_long_GRB}, revealed a highly-obscured, rapidly-fading transient detectable only in the near-infrared. Our late-time observations constrain a luminous rebrightening $\sim$\,$20$ days after the GRB in the rest frame, but deeper observations are needed to rule out rebrightening akin to the relativistic TDE AT2022cmc's slow thermal component \citep{Andreoni2022} or SN~1998bw \citep{Patat2001} given the high redshift and large extinction in the line of sight. 

Using a redshift of $z=1.036$ obtained from JWST observations \citep{Gompertz2025} we modeled the decline of the nIR transient and found it to be consistent with forward shock emission originating from a non-spreading top-hat jet with a Lorentz factor of $\log\Gamma_0=1.25^{+0.16}_{-0.12}$. After correcting for Galactic extinction, we estimated the intrinsic line of sight extinction to be $A_{\textrm{V},\mathrm{GRB}} = 5.67^{+0.3}_{-0.33}$ mag at $z=1.036$. 
While this high extinction is consistent with both the most obscured GRBs \citep{Kruhler2011_dustiest_GRBs} and TDEs \citep{Burrows2011Nat}, the Lorentz factor measurement remains model-dependent: the value derived both without lateral spreading and with lateral spreading ($\Gamma_0 \approx 18$) aligns with relativistic TDEs \citep{Zauderer2011, metzger12_j1644_afterglow, Andreoni2022}, while enforcing additional constraints ($\Gamma_0\theta_\textrm{c}>1$) on a non-spreading jet yields a value ($\Gamma_0 \approx 400$) more typical of GRB collapsars \citep{Liang2010,Ghirlanda2018}. 

Keck, Gemini, and \textit{HST} images enabled the detection of the host galaxy in the $z$, $H$, $K_s$, and $F160W$ filters. \grb's galactocentric offset is consistent with both the population of observed long GRBs and the three known off-nuclear TDEs.
Using the host galaxy SED modeling tools within \textit{Prospector}, we found the galaxy to be massive ($10^{10.66\pm0.15}\,M_\odot$), dusty ($A_{\textrm{V},\mathrm{host}} = 0.88^{+0.47}_{-0.45}$ mag), and highly asymmetric ($A=0.90$ in CAS space). 
The host's inferred stellar mass is consistent with the masses of long GRB hosts and reminiscent of the $\sim 10^{10.9}M_\odot$ host masses observed in all 3 known off-nuclear TDEs \citep{Lin2018, Lin2020, Jin2025, Yao2025ApJ}. The differential between the extinction of the nIR counterpart and the host suggests that \grb\ originated in a dense circumstellar environment or was obscured by a prominent dust cloud within the host. 
The galaxy's high asymmetry makes it an outlier among long GRB hosts and suggests the possibility of a major merger which could support the TDE hypothesis. 

Multiple scenarios remain viable to explain the nature of \grb. The jet modeling highlights the nIR emission accompanying \grb\ as consistent with both a collapsar and a TDE scenarios.  Our data are also consistent with exotic models that could explain the origin of \grb, such as micro-TDEs \citep{Perets2016_micro_TDEs,Beniamini2025}, TDEs from IMBHs \citep{Granot2025}, and helium star merger \citep{Fryer1998,Neights2025}.  The multi-wavelength signatures of the longest-duration \grb\ make it a unique puzzle which may originate from a novel progenitor.

\section*{Acknowledgements}
We thank the anonymous referee for their thoughtful feedback and suggestions that strengthened this work.

We thank Brian Lemaux, staff scientist at Gemini North, for his exceptional assistance throughout this project. His expertise, dedication, and willingness to provide rapid advice across often inconvenient timezones were invaluable to the success of this work. We also thank Chris Simpson for his timely assistance with edge case GMOS reductions. JC, IA, and BO acknowledge useful discussions with Eliza Neights and Eric Burns, and thank Ben Gompertz, Andrew Levan, and Antonio Martin-Carrillo for sharing the redshift derived by \textit{JWST}.

The Andreoni Transient Astronomy Lab is supported by the National Science Foundation award AST 2505775, NASA grant 24-ADAP24-0159, and the Discovery Alliance Catalyst Fellowship Mentors award 2025-62192-CM-19. BO is supported by the McWilliams Postdoctoral Fellowship in the McWilliams Center for Cosmology and Astrophysics at Carnegie Mellon University. MB is supported by a Student Grant from the Wübben Stiftung Wissenschaft. WZ acknowledges the support from the Harvard-Smithsonian Center for Astrophysics through the CfA Fellowship. ERC. acknowledges support from NASA through the Astrophysics Theory Program, grant 80NSSC24K0897. PB's work was funded by a grant (no. 2020747) from the United States-Israel Binational Science Foundation (BSF), Jerusalem, Israel and by a grant (no. 1649/23) from the Israel Science Foundation. PC acknowledges the support from the Zhejiang provincial top-level research support program. CDK gratefully acknowledges support from the NSF through AST-2432037, the HST Guest Observer Program through HST-SNAP-17070 and HST-GO-17706, and from JWST Archival Research through JWST-AR-6241 and JWST-AR-5441. AP is supported by NSF Grant No. 2308193.

This paper includes data gathered with the 6.5 meter Magellan Telescopes located at Las Campanas Observatory, Chile. This paper contains data obtained at the Wendelstein Observatory of the Ludwig-Maximilians University Munich. Funded in part by the Deutsche Forschungsgemeinschaft (DFG, German Research Foundation) under Germany's Excellence Strategy – EXC-2094 – 390783311. This work is based on observations made with
the NASA/ESA Hubble Space Telescope. The data
were obtained from the Mikulski Archive for Space Telescopes at the Space Telescope Science Institute, which
is operated by the Association of Universities for Research in Astronomy, Inc., under NASA contract NAS
5-03127 for JWST. This research is based on data obtained from the Astro Data Archive at NSF NOIRLab. NOIRLab is managed by the Association of Universities for Research in Astronomy (AURA) under a cooperative agreement with the U.S. National Science Foundation. This project used data obtained with the Dark Energy Camera (DECam), which was constructed by the Dark Energy Survey (DES) collaboration. Funding for the DES Projects has been provided by the U.S. Department of Energy, the U.S. National Science Foundation, the Ministry of Science and Education of Spain, the Science and Technology Facilities Council of the United Kingdom, the Higher Education Funding Council for England, the National Center for Supercomputing Applications at the University of Illinois at Urbana-Champaign, the Kavli Institute of Cosmological Physics at the University of Chicago, Center for Cosmology and Astro-Particle Physics at the Ohio State University, the Mitchell Institute for Fundamental Physics and Astronomy at Texas A\&M University, Financiadora de Estudos e Projetos, Fundacao Carlos Chagas Filho de Amparo, Financiadora de Estudos e Projetos, Fundacao Carlos Chagas Filho de Amparo a Pesquisa do Estado do Rio de Janeiro, Conselho Nacional de Desenvolvimento Cientifico e Tecnologico and the Ministerio da Ciencia, Tecnologia e Inovacao, the Deutsche Forschungsgemeinschaft and the Collaborating Institutions in the Dark Energy Survey. 

Some of the data presented herein were obtained at the W. M. Keck Observatory, which is operated as a scientific partnership among the California Institute of Technology, the University of California and the National Aeronautics and Space Administration. The Observatory was made possible by the generous financial support of the W. M. Keck Foundation. The authors wish to recognize and acknowledge the very significant cultural role and reverence that the summit of Maunakea has always had within the indigenous Hawaiian community. 
Based on observations at Cerro Tololo Inter-American Observatory, NSF’s NOIRLab (NOIRLab Prop. ID 2025A-729671; PI: Palmese), which is managed by the Association of Universities for Research in Astronomy (AURA) under a cooperative agreement with the National Science Foundation. Based on observations obtained at the international Gemini Observatory, a program of NSF's OIR Lab, which is managed by the Association of Universities for Research in Astronomy (AURA) under a cooperative agreement with the National Science Foundation on behalf of the Gemini Observatory partnership: the National Science Foundation (United States), National Research Council (Canada), Agencia Nacional de Investigaci\'{o}n y Desarrollo (Chile), Ministerio de Ciencia, Tecnolog\'{i}a e Innovaci\'{o}n (Argentina), Minist\'{e}rio da Ci\^{e}ncia, Tecnologia, Inova\c{c}\~{o}es e Comunica\c{c}\~{o}es (Brazil), and Korea Astronomy and Space Science Institute (Republic of Korea). The data were acquired through the Gemini Observatory Archive at NSF NOIRLab and processed using DRAGONS (Data Reduction for Astronomy from Gemini Observatory North and South).

The Collaborating Institutions are Argonne National Laboratory, the University of California at Santa Cruz, the University of Cambridge, Centro de Investigaciones Energeticas, Medioambientales y Tecnologicas-Madrid, the University of Chicago, University College London, the DES-Brazil Consortium, the University of Edinburgh, the Eidgenossische Technische Hochschule (ETH) Zurich, Fermi National Accelerator Laboratory, the University of Illinois at Urbana-Champaign, the Institut de Ciencies de l'Espai (IEEC/CSIC), the Institut de Fisica d'Altes Energies, Lawrence Berkeley National Laboratory, the Ludwig Maximilians Universitat Munchen and the associated Excellence Cluster Universe, the University of Michigan, NSF NOIRLab, the University of Nottingham, the Ohio State University, the University of Pennsylvania, the University of Portsmouth, SLAC National Accelerator Laboratory, Stanford University, the University of Sussex, and Texas A\&M University.

The HST data presented in this article were obtained from the Mikulski Archive for Space Telescopes (MAST) at the Space Telescope Science Institute. The specific observations analyzed can be accessed via \dataset[doi: 10.17909/12kv-9c88]{https://doi.org/10.17909/12kv-9c88}.

\software{\texttt{Astropy} \citep{Astropy2013, Astropy2018, Astropy2022}, \texttt{matplotlib} \citep{Hunter2007}, \texttt{SWarp} \citep{Bertin2010}, \texttt{Source Extractor} \citep{Bertin1996}, \texttt{Morfometryka} \citep{Ferrari2015}, \texttt{GALFIT} \citep{Peng2002}, \texttt{DRAGONS} \citep{Labrie2019}, \texttt{Photutils} \citep{Bradley_2024}, \texttt{VegasAfterglow} \citep{2025arXiv250710829W}, \texttt{astrometry.net} \citep{Lang2010}, \texttt{dynesty} \citep{dynesty}, \texttt{DrizzlePac}, \texttt{emcee} \citep{emcee}, \texttt{bilby} \citep{2019ApJS..241...27A}, \texttt{Claude} \citep{Claude_AI}}

\bibliographystyle{aasjournal}
\bibliography{references}

\appendix

\section{Observation Details}
\label{sec:observations_long}

\subsection{DECam}
We observed the location of \grb\ with the Dark Energy Camera \citep[DECam;][]{Flaugher2015} mounted on the Victor M. Blanco 4-meter Telescope at Cerro Tololo Inter-American Observatory (Program ID 2025A-729671; PI: Palmese), beginning approximately 15.3 hours after the first GBM detection.
Observations spanned from 2025-07-03 05:12:58 UT to 2025-07-03 05:27:50 UT, and consisted of exposures in 4 filters: $1 \times 70$ s in $g$, $1 \times 80$ s in $r$, $1 \times 90$ s in $i$, and $2 \times 100$ s in $z$-band. 
The images were reduced using the DESDM pipeline V5.5.3 \citep{decam_pipeline} with astrometric and photometric calibration against United States Naval Observatory B catalog \citep{Monet2003}. No optical counterpart or host galaxy was detected in any filter, with $5\sigma$ limiting magnitudes of $g=23.56$, $r=23.04$, $i=22.26$, and $z=22.21, 22.08$ AB mag for the two $z$-band exposures, respectively.

\subsection{Fruanhofer Telescope at Wendelstein Observatory}
\label{sec:wendelstein}
We observed the location of \grb\ with the Fraunhofer Telescope at Wendelstein Observatory (FTW) using the Three Channel Imager (3KK, \citealt{3kk_spie}) for two epochs. The first epoch began at 2025-07-03 21:52:47 UT ($\sim32$ hours after the first GBM detection) and simultaneously imaged in the $r$, $i$, and $J$ bands for $40 \times 180$ s. No sources were detected within the {\it Swift}/XRT localization down to $5\sigma$ limiting magnitudes of $r=22.47$, $i=22.94$, $J=21.25$ AB mag.

A second epoch was obtained at 2025-07-05 10:14:00 UT ($\sim68$ hours after the first GBM detection) with simultaneous imaging in the $r$, $i$, and $K$ bands for $101 \times 60$ s. Using later imaging from Magellan (Sec. \ref{sec:magellan}) as a template for image subtraction we detect the nIR counterpart at $K=21.07\pm0.16$ AB mag. No detections were made in the optical bands down to $5\sigma$ limiting magnitudes of $r=22.06$ and $i=21.79$ mag. For both epochs we use the Pan-STARRS DR1 catalog \citep{Pan-STARRS1} to photometrically calibrate $r$ and $i$ bands, and use the 2MASS catalog \citep{2mass_catalog} to photometrically calibrate $J$ and $K$. 
We applied standard Vega to AB conversions \citep{Blanton2007} to convert the calibrated $J$ and $K$ magnitudes to the AB system.

\subsection{Gemini Telescopes}
\subsubsection{Gemini South}
We observed the location of \grb\ in $z$-band using the Gemini Multi Object Spectrograph on Gemini South (GMOS-S; program ID GS-2025A-Q-117; PI: Andreoni) in imaging mode beginning at 2025-07-04T03:43:54 UT ($\sim38$ hours after the first GBM detection). We obtained $120 \times 60$ s exposures for a total integration time of 2.0 hours. The images were reduced using the \texttt{DRAGONS} pipeline \citep{DRAGONS} and photometric calibrations were obtained with the Pan-STARRS DR1 (PS1) catalog \citep{Pan-STARRS1}. No transient counterpart was detected to a $5\sigma$ limiting magnitude of $z = 23.95$ AB mag. 

\subsubsection{Gemini North}
We observed the location of \grb\ in $z$-band using the Gemini Multi Object Spectrograph on Gemini North (GMOS-N; program ID GN-2025A-Q-213; PI: Andreoni) in imaging mode beginning at 2025-07-20T09:55:45 UT ($\sim18$ days after the first GBM detection). We obtained $52 \times 120$ s exposures for a total integration time of 1.73 hours. We did not coadd with the first epoch due to differences in readout mode. The images were reduced using the \texttt{DRAGONS} pipeline \citep{DRAGONS} and photometric calibrations were obtained with the PS1 catalog. No counterpart was detected to a $5\sigma$ limiting magnitude of $z = 24.19$ AB mag. We detected extended emission from the host galaxy using a 2.25\arcsec $\times$ 1.0\arcsec\ elliptical aperture, measuring $z=23.20\pm0.13$ mag (Figure \ref{fig:hostfig}).

\subsection{Keck}
\label{sec:keck}
We observed the counterpart to \grb\ with the Multi-Object Spectrograph for Infrared Exploration \citep[MOSFIRE;][]{mosfire_spie} on the W. M. Keck Observatory telescope I (PI: Kasliwal, Program ID: C348), obtaining $J$, $H$, and $K_s$-band images of the afterglow. Observations began at 2025-07-05 10:05:00 UT ($\sim68$ hours after the first GBM detection) and consisted of two box-9 dither sequences: 8 s exposures with 4 coadds in $K_s$, 4 s exposures with 8 coadds in $H$, and one box-9 dither of 11 s exposures with 3 coadds in $J$. The data were reduced using standard procedures, and photometric calibration was performed against the VISTA Hemisphere Survey \citep[VHS;][]{McMahon13} DR4.1 catalog for $K_s$ and $J$ bands and photometrically calibrated using the UKIRT Infrared Deep Sky Survey (UKIDDS; \cite{Lawrence07}) catalog in $H$-band.
We clearly detect the host galaxy in the $K_s$ and $H$-bands: $K_s=20.50\pm0.07$ AB mag (using a 2.88\arcsec $\times$ 1.08\arcsec\ aperture) and $H=21.55\pm0.17$ AB mag (using 2.26\arcsec $\times$ 0.9\arcsec\ aperture. We perform image subtraction using $K_s$ images taken $\sim 38$ days later with the Magellan telescope (Sec. \ref{sec:magellan}) and measure the transient to be $K_s=21.96$ AB mag.

\subsection{NEWFIRM}
\label{sec:newfirm}
We observed the location of \grb\ at two epochs with the recommissioned NOAO Extremely Wide Field Infrared Imager (NEWFIRM; \citealt{Autry03}) on the Victor M. Blanco 4-meter Telescope at Cerro Tololo Inter-American Observatory. 
The first epoch began at 2025-07-10 19:26:50 UT ($\sim8$ days after the first GBM detection) and consisted of $51 \times 60$\,s exposures in $H$ band. No transient emission was detected down to a $5\sigma$ limiting magnitude of $H=20.50$ AB mag. 
The second epoch began the following night at 2025-07-11 21:33:30 UT ($\sim9$ days after the first GBM detection) with an additional $32 \times 60$\,s exposures in $H$ band, followed by $120 \times 60$ s exposures in $K_s$ band beginning at 2025-07-11 21:50:51 UT. All data were processed using the \texttt{photpipe} imaging and photometry pipeline \citep{Rest05}. Due to poor seeing conditions and high airmass during both nights, the combined $H$-band dataset achieved only a $5\sigma$ limiting magnitude of $H=20.10$ AB mag. The $K_s$-band observations reached $K_s=20.62$ AB mag ($5\sigma$). No transient was detected in either band, even after image subtraction with a later $K_s$ epoch (Sec. \ref{sec:magellan}). We detected the host galaxy in $K_s$-band at $K_s=21.34\pm0.48$ AB mag using a $2.4\arcsec \times 1.2$\arcsec\ elliptical aperture. Photometric calibration was performed using VHS DR4.1 \citep{McMahon13} for $K_s$-band and UKIDSS \citep{Lawrence07} for $H$-band due to limited VHS $H$-band coverage in this field.

\subsection{Magellan}
\label{sec:magellan}
We obtained one epoch of $K_s$ band imaging with the near-infrared imager FourStar \citep{Persson2013} mounted on the 6.5 m Magellan Baade Telescope. The observations were conducted at 2025-08-12 06:00:00 UT ($\sim41$ days post burst) by executing a sequence of exposures at 12 random dithered positions, with each position consisting of 12 loops of 5.822 second integration. The total exposure time on target was 1397.28\,s. The FourStar images were reduced using custom Python scripts following the standard procedures: dark subtraction, flat-fielding, sky background subtraction, astrometric registration of the science images using \texttt{astrometry.net} \citep{Lang2010}, and image stacking with \texttt{SWarp} \citep{Bertin2002}.
No transient emission was detected down to a $5\sigma$ limiting magnitude of $K_s = 21.94$ AB mag. The host galaxy was detected at $K_s=21.32\pm0.21$ AB mag using a $2.4\arcsec \times 1.12\arcsec$ elliptical aperture (Figure \ref{fig:hostfig}). This template was used as a reference for image subtraction for the $K_s$ images discussed in sections \ref{sec:wendelstein}, \ref{sec:keck}, and \ref{sec:newfirm}.

\subsection{Hubble Space Telescope}
\label{HST}

We reprocessed publicly available \textit{Hubble Space Telescope} (\textit{HST}) imaging obtained from the Mikulski Archive for Space Telescopes (MAST)\footnote{\url{https://archive.stsci.edu/index.html}}. The \textit{HST} data (Program ID 17988; PI: Levan) were previously published in \citet{Levan_long_GRB}. The images were obtained with the Wide Field Camera 3 (WFC3) in the $F160W$ filter at 2025-07-15 02:39:23 ($\sim12.5$ days after the first GBM detection). The data were reduced using standard procedures within the \texttt{DrizzlePac} software package. We reprocessed the data to a final pixel scale of $0.06\arcsec$/pix using \texttt{pixfrac}\,$=$\,$0.8$. Using a 1.75\arcsec $\times$ 0.65\arcsec\ elliptical aperture we measure a host magnitude with transient contamination of $F160W=21.50\pm0.1$ mag (Figure \ref{fig:hostfig}), where the error bars have been increased to include systematics. In Section \ref{sec:host_morph} we model the host morphology as either a spiral galaxy or as two galaxies undergoing a merger. We subtracted both of these models from the reprocessed image and performed photometry on the residuals (Figure \ref{fig:host_morphology}). 
Using a $0.30$\arcsec\ diameter circular aperture we measure the transient magnitude to be $F160W=26.52\pm0.1$ AB mag in the residual of the spiral galaxy model and $F160W=26.75\pm0.1$ AB mag in the residual of the merger model, again with our error bars increased to include systematics. The residual flux after subtraction using the galaxy merger model is more point spread function (PSF) like, so we employ a magnitude of $F160W=26.52\pm0.1$ AB mag in our model fitting.

\section{Host photometry}
Host photometry presented in this work are displayed in Table \ref{tab:host_observations_table}.

\begin{table*}[htbp]
    \centering
        \caption{Host galaxy photometry of \grb. All magnitudes are provided in AB system and are not corrected for Galactic extinction. The aperture column denotes the semimajor $\times$ semiminor axis of the elliptical aperture used for forced photometry.}
    \label{tab:host_observations_table}
    \begin{tabular}{c|c|c|c|c}
    \hline
    \hline
         Datetime & Instrument & Filter & Host Mag & Aperture\\
\hline         
2025-07-05 10:14:00 & Keck/MOSFIRE & $K_s$ & 20.50 ± 0.07 & 2.88$''$ $\times$ 1.08$''$\\
2025-07-11 21:50:51 & NEWFIRM & $K_s$ & 21.34 ± 0.48 & 2.4$''$ $\times$ 1.2$''$ \\
2025-08-12 06:00:00 & Magellan/FourStar & $K_s$ & 21.32 ± 0.21 & 2.4$''$ $\times$ 1.12$''$ \\
\hline
2025-07-05 10:14:00 & Keck/MOSFIRE & $H$ & 21.55 ± 0.17 & 2.26$''$ $\times$ 0.9$''$ \\
2025-07-15 02:39:23 & HST/WFC3 & $F160W$ & 21.50 ± 0.1 & 1.75$''$ $\times$ 0.65$''$ \\
\hline
2025-07-20 09:55:45 & Gemini/GMOS-N & $z$ & 23.20 ± 0.13 & 2.25$''$ $\times$ 1.0$''$\\
\hline
    \end{tabular}
\end{table*}

\section{Additional Model Fitting Results}

\begin{figure*}
    \centering
\includegraphics[width=\linewidth]{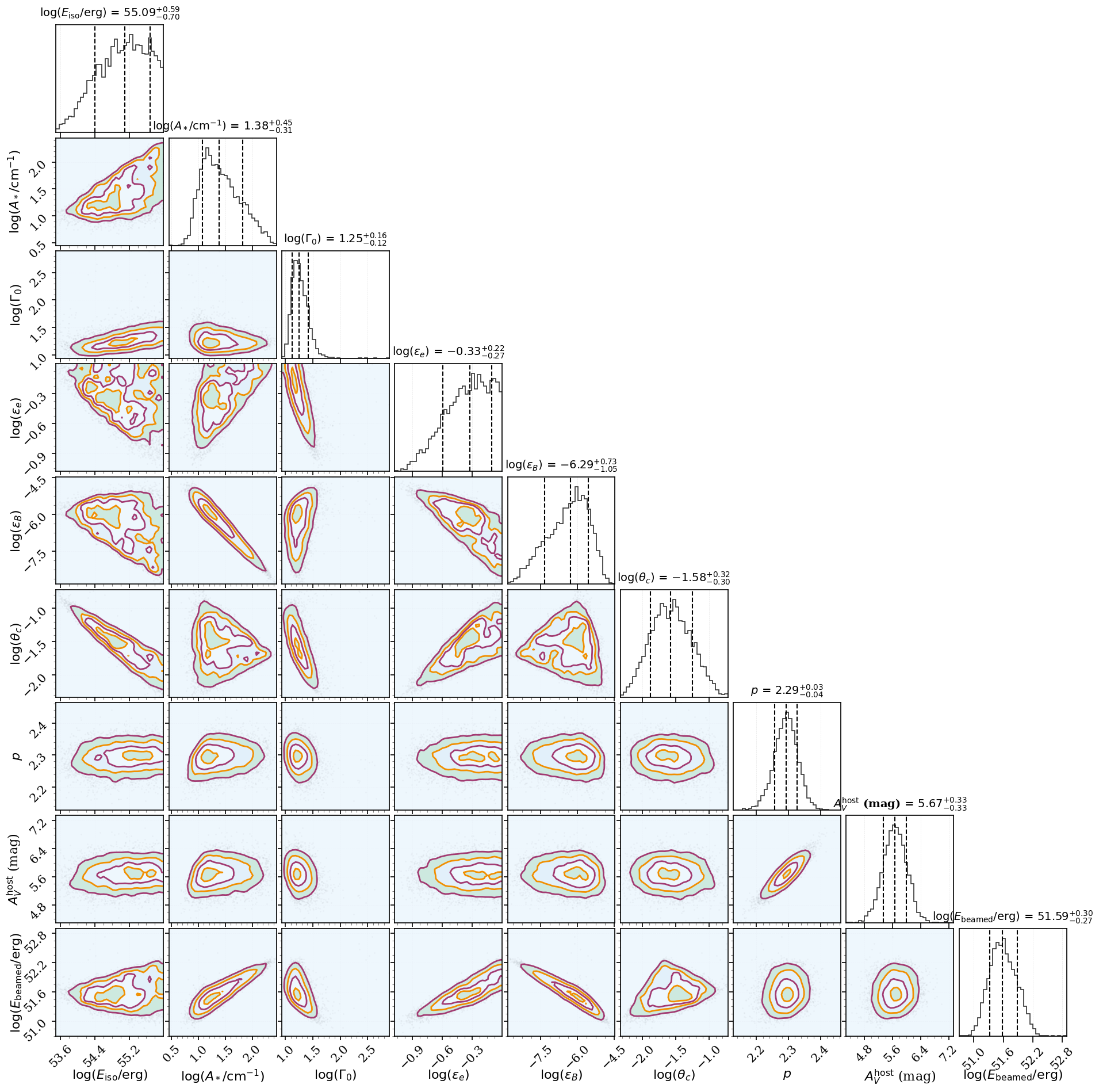}  
    \caption{
    Posterior distributions of the jet parameters derived using \texttt{VegasAfterglow} to fit the available multi-wavelength data with a non-spreading top-hat jet. }
\label{fig:jet_corner_plot-nospread}
\end{figure*}

\begin{figure*}
    \centering
\includegraphics[width=\linewidth]{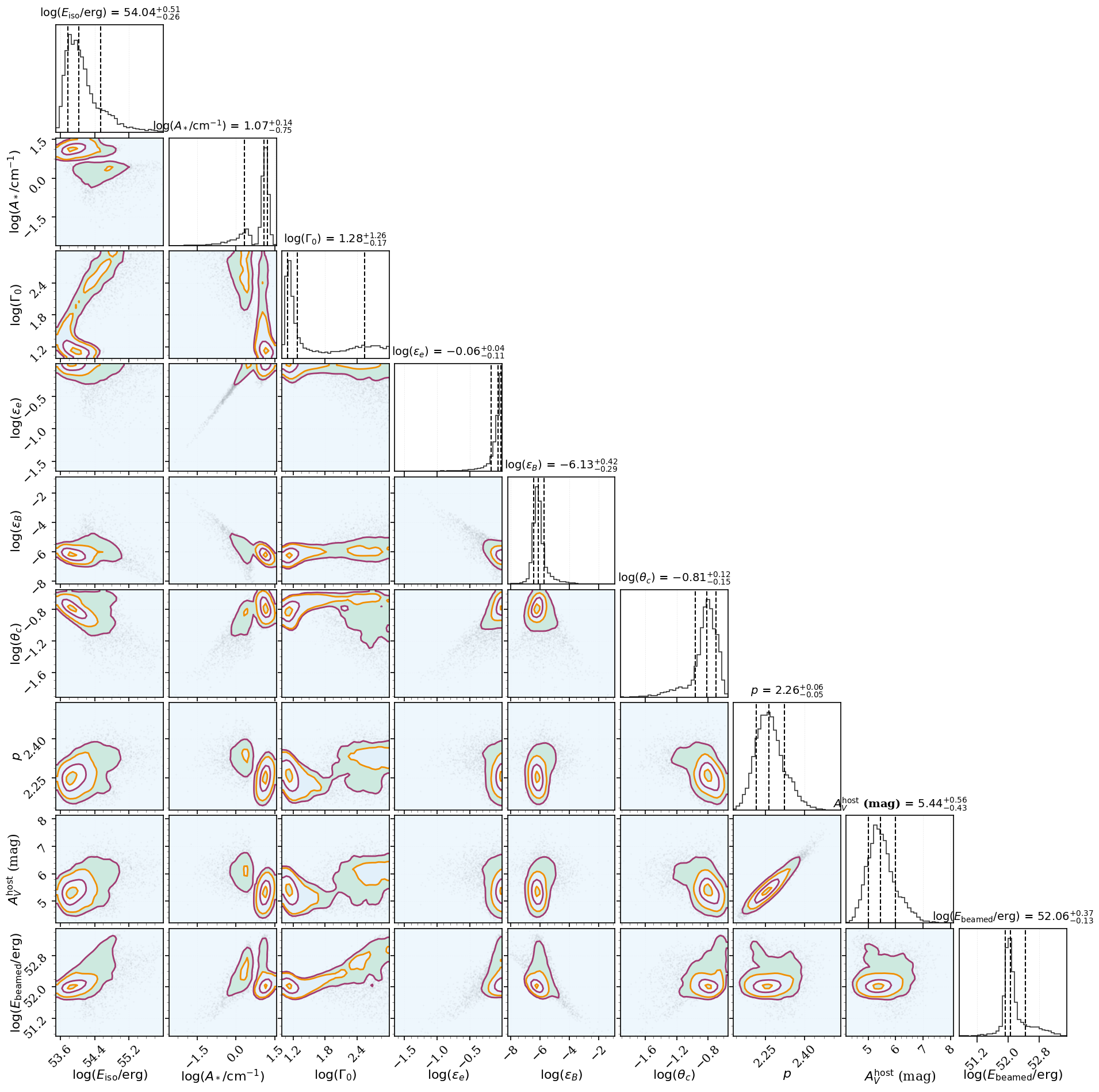}  
    \caption{
    Posterior distributions of the jet parameters derived using \texttt{VegasAfterglow} to fit the available multi-wavelength data with the inclusion of lateral jet spreading effects.  }
\label{fig:jet_corner_plot-spread}
\end{figure*}

\begin{figure*}
    \centering
\includegraphics[width=\linewidth]{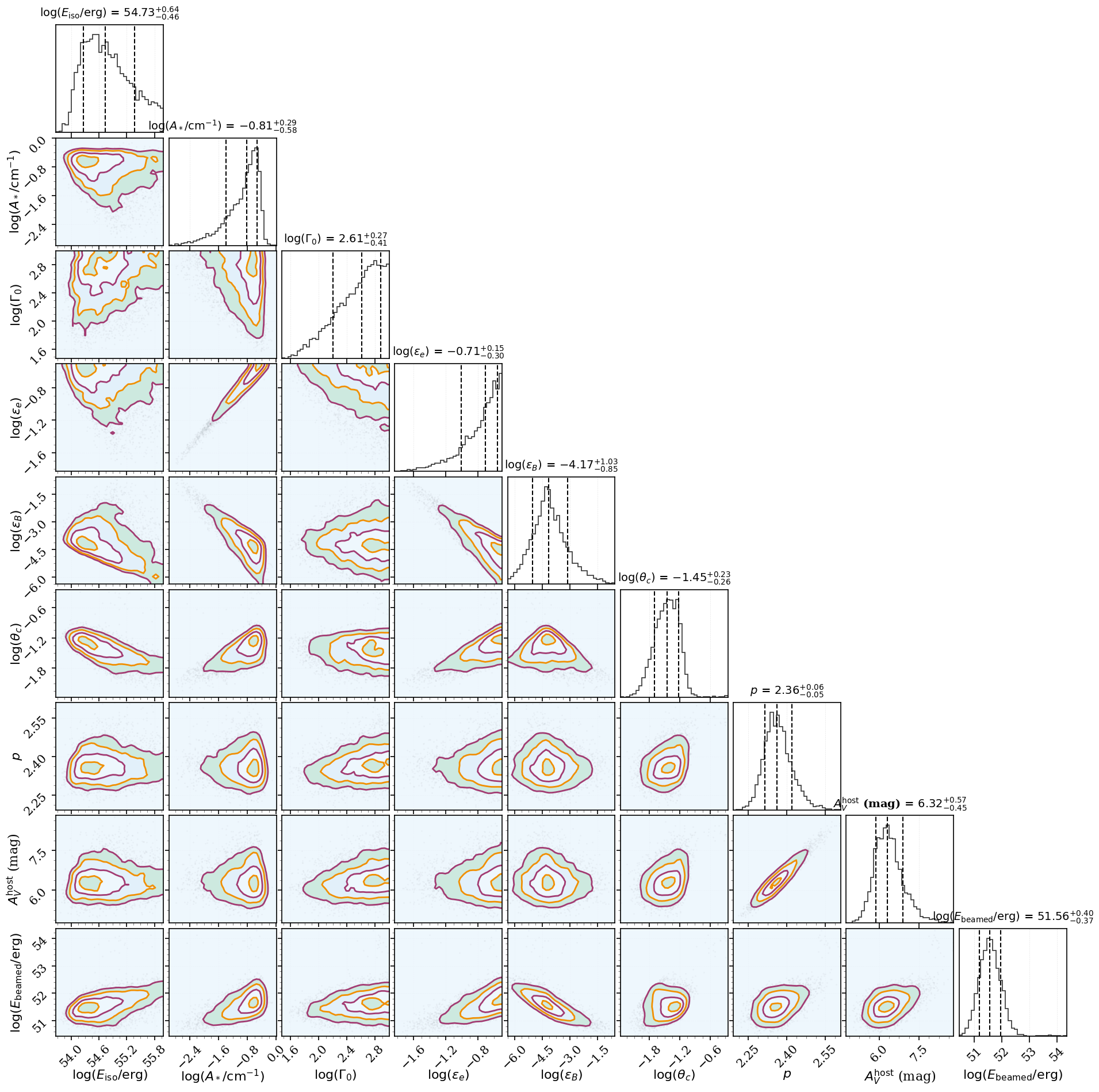}  
    \caption{
    Posterior distributions of the jet parameters derived using \texttt{VegasAfterglow} to fit the available multi-wavelength with a non-spreading top-hat jet  constrained such that $\Gamma_0\theta_\textrm{c}>1$.  }
\label{fig:jet_corner_plot-constraint}
\end{figure*}

\begin{figure*}
    \centering
\includegraphics[width=\linewidth]{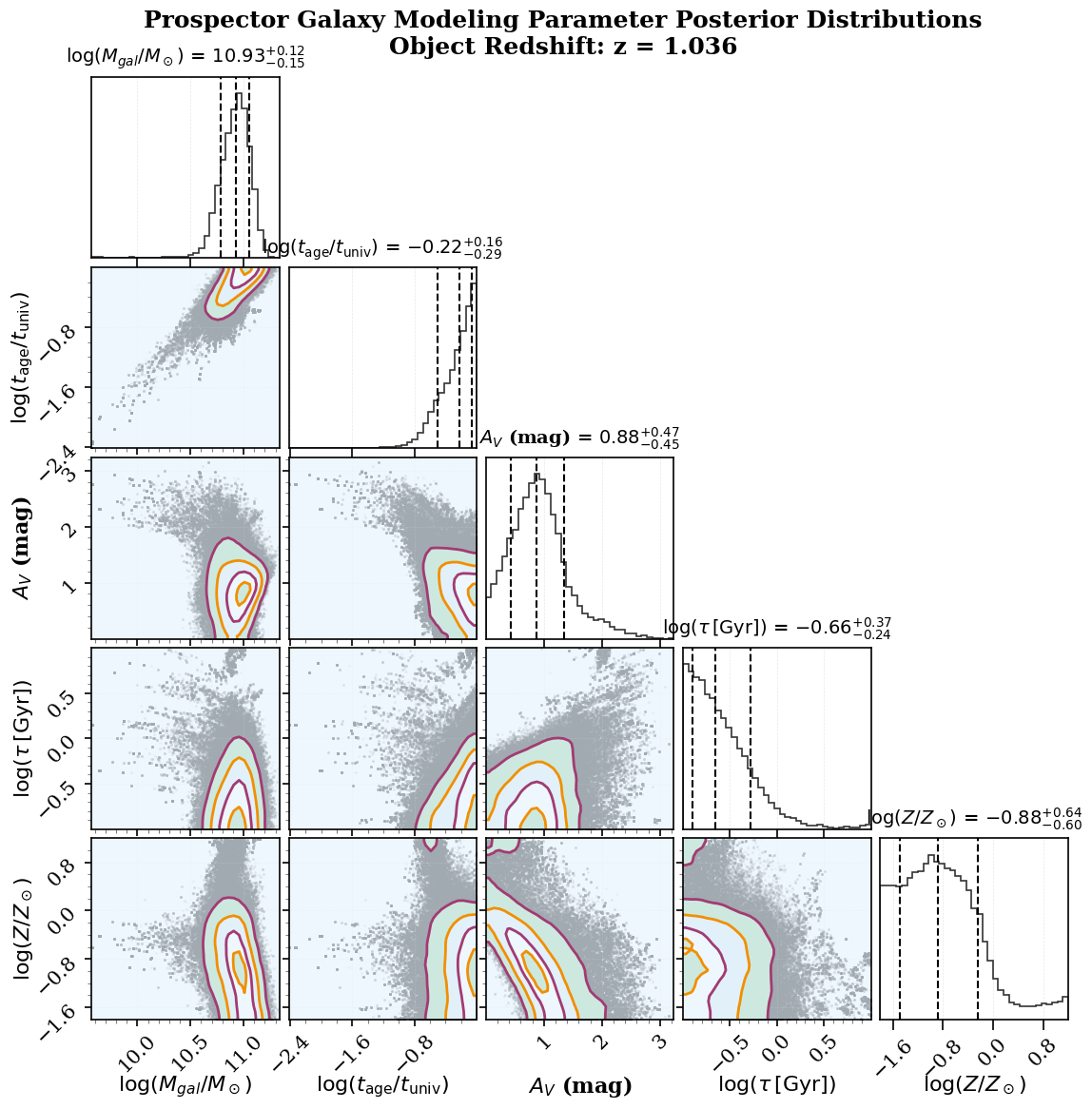}  
    \caption{Posterior distributions of our host galaxy SED fitting results using \texttt{prospector}. We note that $M_\textrm{gal}$ refers to the total mass formed, and not the galaxy's stellar mass. See also Figure \ref{fig:prospector}. }
\label{fig:prospector_corner_plot}
\end{figure*}

The results of our jet modeling are displayed in Figures \ref{fig:jet_corner_plot-nospread}, \ref{fig:jet_corner_plot-spread}, and \ref{fig:jet_corner_plot-constraint}. The results of our host galaxy SED fitting are displayed in Figure \ref{fig:prospector_corner_plot}.

\end{document}